\documentclass[printer]{aa}
\usepackage{txfonts}
\usepackage{graphicx}
\usepackage{graphics}
\usepackage{natbib}
\usepackage{float}
\usepackage{placeins}
\usepackage{caption}
\usepackage{multirow}
\usepackage{lscape}
\bibpunct{(}{)}{;}{a}{}{,} 

\begin{document}
\newcommand{\msun}{\mbox{M$_{\odot}$}}
\newcommand{\rsun}{\mbox{R$_{\odot}$}}
\newcommand{\lsun}{\mbox{L$_{\odot}$}}
\newcommand{\tsun}{\mbox{T$_{\odot}$}}
\title{Fundamental parameters of four massive eclipsing binaries in Westerlund~1}

\author{E. Koumpia\inst{\ref{inst1},\ref{inst3}} \and A.Z. Bonanos\inst{\ref{inst2},\ref{inst3}}}

\institute{SRON Netherlands Institute for Space Research, Landleven 12,
  9747 AD Groningen, The Netherlands; Kapteyn Institute, University of
  Groningen, The Netherlands\label{inst1}
\and 
National Observatory of Athens, Institute of Astronomy, Astrophysics, Space Applications \& Remote Sensing, I. Metaxa \& Vas. Pavlou St., Palaia Penteli, GR-15236 Athens,
Greece\label{inst2}
\and \email{e.koumpia@sron.nl;bonanos@astro.noa.gr}\label{inst3}
}
\date{Received date/Accepted date}

\begin{abstract}
{Only a small number of high mass stars ($>30\,\emph{\msun}$) have fundamental parameters (i.e. masses and radii) measured with high enough accuracy from eclipsing binaries to constrain formation and evolutionary models of massive stars.} {This work aims to increase this limited sample, by studying the four massive eclipsing binary
  candidates discovered by Bonanos in the young massive cluster Westerlund~1.}  {We present
  new follow-up echelle spectroscopy of these binaries and models of their light and radial
  velocity curves.} {We obtain fundamental parameters for
  the eight component stars, finding masses that span a range of
  $10-40\,$\emph{\msun}, and contributing accurate fundamental parameters 
for one additional very massive star, the 33 $\emph{\msun}$ component of W13. WR77o is found to have a $\sim40$ $\emph{\msun}$ companion, which provides a second dynamical constraint on the mass of the progenitor of the magnetar known in the cluster. We
  also use W13 to estimate the first, direct, eclipsing binary distance to Westerlund~1 and therefore the magnetar and find it to be at $3.7\pm0.6$~kpc.}
{Our results confirm previous evidence for a high mass for the progenitor of the magnetar. In addition, the availability of eclipsing binaries with accurate parameters opens the way for direct, independent, high precision eclipsing binary distance measurements to Westerlund~1.}

\end{abstract}

\keywords{stars: fundamental parameters, stars: massive, binaries:
  eclipsing, stars: early-type, stars: Wolf-Rayet, open clusters and
  associations: individual (Westerlund~1)} 

\titlerunning{Wd 1 Eclipsing Binaries} 
\authorrunning{Koumpia \& Bonanos} 
\maketitle


\section{Introduction} 

Improving stellar structure, formation and evolution models of massive
stars is limited by the accuracy with which their fundamental parameters
(masses and radii) are known. \citet{Andersen91} reviewed the importance
of accurate parameters from eclipsing binaries, while \citet{Torres10}
revisited the subject, reviewing the advances made possible by the
continuous improvements in observational accuracy. They also compiled a catalog of
detached systems with measurements accurate to $\pm3\%$ from the
literature, which consists of 95 systems and extends the mass range from
$0.2\,\emph{\msun}$ to $27\,\emph{\msun}$. The earliest-type star in the sample was an
O7 star with a mass of $27.3\,\emph{\msun}$. \citet{Bonanos09} found that the
less stringent accuracy criterion of $10\%$ is only fulfilled by 15
eclipsing binaries with masses above $30\,\emph{\msun}$, this number having only increased
to 17 \citep[with the components of Cyg OB2-17, measured by][]{Stroud10}
in the last three years. This astounding lack of accurate measurements of
the fundamental parameters of the highest mass stars, combined with the
large parameter space of metallicity and evolutionary stage that they
span, which is further complicated by multiplicity and in particular binary
evolution, serves as the motivation for this work.

To improve the situation, a systematic survey of the most massive
stars in the Local Group has been undertaken \citep[see][for an
  overview]{Bonanos10}. As part of this broader survey,
\citet{Bonanos07} conducted the first variability survey of the young
massive ``super star cluster'' Westerlund~1 
\citep{Clark05} in search of
massive eclipsing binary candidates, as it is one of the most massive young
clusters known in the Local Group, with an age of 4.5-5 Myr
\citep{Crowther06}. Wd~1 is of great interest, as it contains an
assortment of rare evolved massive stars, such as blue, yellow and red
supergiants, including a rare supergiant B[e] star \citep{Clark05,
  Negueruela10}, 24 confirmed Wolf-Rayet stars \citep{Crowther06}, a
luminous blue variable \citep{Clar04}, and a magnetar \citep{Muno06},
thereby making possible the determination of fundamental parameters of
massive stars at different evolutionary stages. The photometric
variability search resulted in four massive eclipsing binary candidates
\citep[W$_{DEB}$, W13, W36, and WR77o;][]{Bonanos07}, which are the subject
of this work.

In the following sections, we present follow-up spectroscopy and modeling of these four
binaries, and the resulting component parameters. We present the observations in Section 2, the radial velocity measurements in Sect. 3, the models for each eclipsing binary in Sect. 4, the comparison with evolutionary models in Sect. 5, an estimation of the distance in Sect. 6, and a discussion of our results and conclusions in Sect. 7.

\clearpage
\section{Observations and data reduction} 

Multi-epoch spectroscopy of the four eclipsing binaries was obtained
during seven nights in 2007-2008 and one night in 2011 using the MIKE
spectrograph \citep{Bernstein03} at the 6.5 meter Magellan (Clay)
telescope at Las Campanas Observatory, Chile. This camera uses a
$2048\times4096$ SITe CCD with a pixel size of $15\,\mu m$ pixel$^{-1}$ and
a pixel scale of $0\farcs13$ pixel$^{-1}$. The $1.0\arcsec\times
5.0\arcsec$ slit yielded a spectral resolution $R \sim 24\,000$
(12~km~s$^{-1}$; the exact value depending on the binning used) at 7000$\AA$, as measured from the full width at half
maximum of the comparison lamp lines. Additional spectra were obtained over four consecutive nights in 2007 with
the Inamori--Magellan Areal Camera \& Spectrograph (IMACS) Multi-Object Echelle (MOE) spectrograph \citep{Dressler11} on the 6.5 meter Magellan
(Baade) telescope. The 0.6 $\arcsec$ slit yielded a resolving 
power of $R=21\,000$ (14~km~s$^{-1}$).

The IMACS spectra were reduced and extracted with the Carnegie Observatories System for MultiObject Spectroscopy (COSMOS) pipeline
(version 2.15), while the MIKE spectra with the MIKE python pipeline
(version 2.5.4), written by D. Kelson \citep{Kelson00, Kelson03}. The
extracted orders for each star were averaged, normalized, and merged,
yielding a wavelength range 5450$\AA$$-$9400$\AA$ for the MIKE and 6380$\AA$$-$8750$\AA$ 
for the IMACS spectra, 
since the large amount of 
reddening toward Westerlund~1 \citep[$A_{V}\sim11$~mag;][]{Negueruela10} 
limited the signal-to-noise ratio (S/N) at shorter wavelengths. Heliocentric radial velocity 
corrections for each star were computed with the IRAF\footnote {IRAF is distributed by the National Optical Astronomy Observatory, which is operated by the Association of Universities for Research in Astronomy, Inc., under cooperative agreement with the NSF.} $rvsao.bcvcorr$ routine and were taken into account in the subsequent radial velocity
determination. The spectra of W$_{DEB}$ and W36 displayed double hydrogen and helium absorption lines consistent with a late-O or
early-B spectral type, W13 displayed hydrogen and helium lines, in both absorption and emission, 
while WR77o displayed strong emission
in \ion{He}{II}~$\lambda\lambda6562,8237$ and
\ion{N}{IV}~$\lambda\lambda7103-7128$, consistent with the WN7o spectral
type determined by \citet{Crowther06}. Even though the S/N was not sufficient to detect the
companion, we proceeded with the analysis of this single lined
binary.

\begin{table}[h]
\caption{Log of Magellan spectroscopic observations}

\centering
\small\addtolength{\tabcolsep}{-4.7pt}
\begin{tabular}{c c c c c c c c }
\hline\hline
UT Date & Instru- & Resolu- & Disper- & \multicolumn{4}{c}{Total Exposure Time (sec) / S/N} \\ & ment & tion & sion & & & & \\ 
 & & ($\lambda/\Delta\lambda$) & ($\AA/$pix) & W$_{DEB}$ & W36 & W13 & WR77o \\
\hline\hline

$01/03/2007$ & MIKE & 25500 & 0.05 & \ldots  & 1800/14 &  1440/30 &  4800/7 \\
$08/03/2007$ & MIKE & 24200 & 0.10 & 3600/22 &  2700/38 &  1440/50 &  4800/12 \\
$09/03/2007$ & MIKE & 24200 & 0.10 & 3600/20 & \ldots &  1440/55  & 3300/14 \\
$15/03/2007$ & MIKE & 24200 & 0.10 & \ldots &  2700/42 &  1260/49 &  3600/14 \\
$28/06/2007$ & MOE & 21000 & 0.25 & \ldots  & \ldots  & 1800/31  & \ldots  \\
$29/06/2007$ & MOE & 21000 & 0.25 & \ldots  & 2400/20  & 2400/35  & \ldots  \\
$30/06/2007$ & MOE & 21000 & 0.25 & \ldots  & \ldots  & 3600/17  & \ldots  \\
$01/07/2007$ & MOE & 21000 & 0.25 & \ldots  & 2400/35  & \ldots  & 2400/7  \\
$03/06/2008$ & MIKE & 24000 & 0.10 & 4800/13 &  900/14  & 1800/28 &  4800/7 \\
$30/08/2008$ & MIKE & 23800 & 0.15 & \ldots & 2700/56  & 1440/81 &  3600/27 \\
$01/09/2008$ & MIKE & 23800 & 0.15 & \ldots & 2700/54 &  1440/80 &  3600/20 \\
$01/09/2011$ & MIKE & 24200 & 0.10 & 3600/32  & \ldots &  \ldots & \ldots \\
\hline
\end{tabular}
\label{obslog}
\end{table}

Table~\ref{obslog} lists the UT
dates of the observations, the spectrograph and the corresponding 
resolution, dispersion
and total exposure time and S/N for each target, determined at 7000 $\AA$. Typically three exposures of
each target were obtained with the MIKE spectrograph to enable cosmic ray 
removal. The empty records indicate spectra with very low S/N, which made the detection of the lines and thus the use of those spectra impossible.

\section{Radial velocity analysis} 

\begin{figure*}[h]
\begin{center}$
\begin{array}{cc}
\includegraphics[width=2.5in, angle=90]{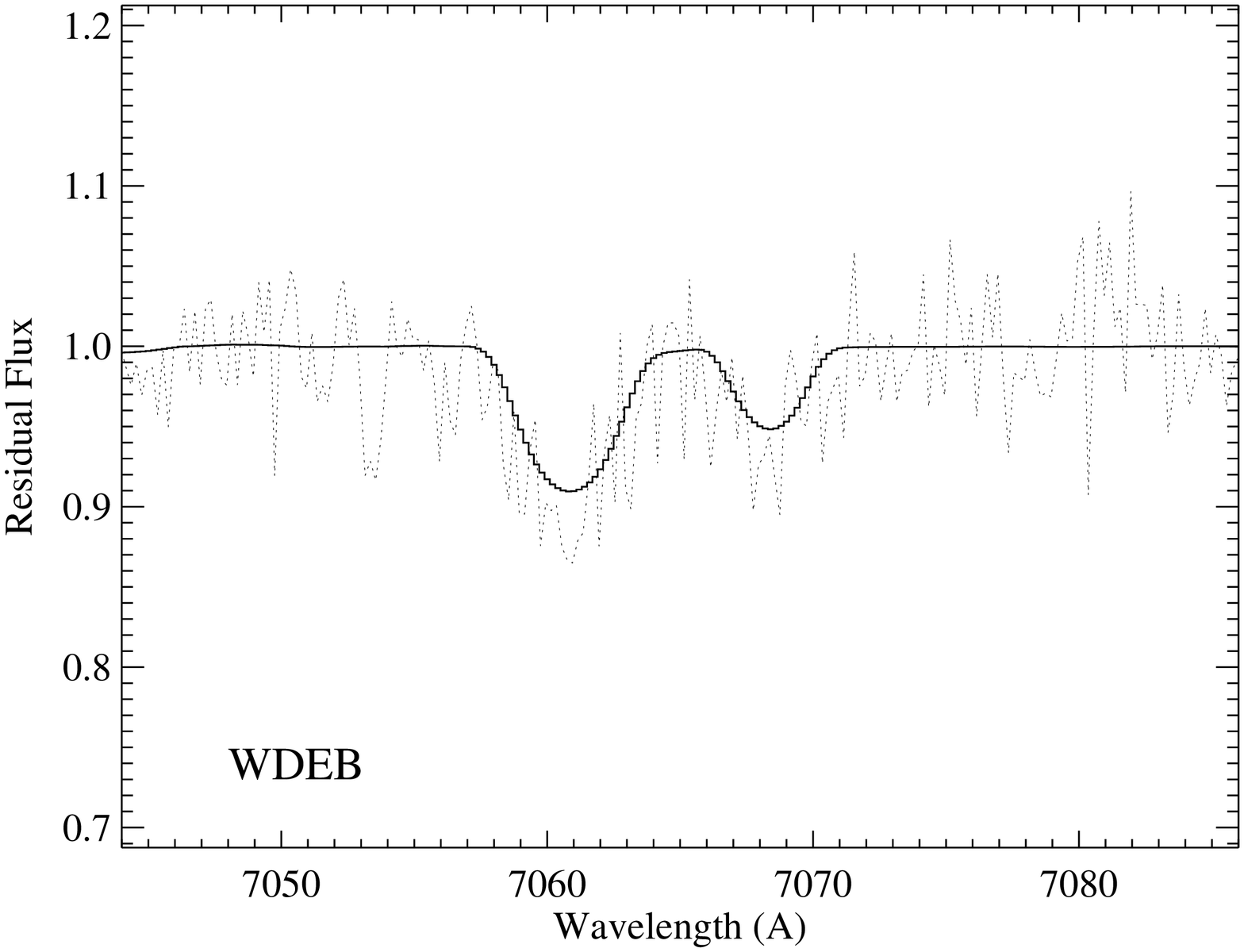} &
\includegraphics[width=2.5in, angle=90]{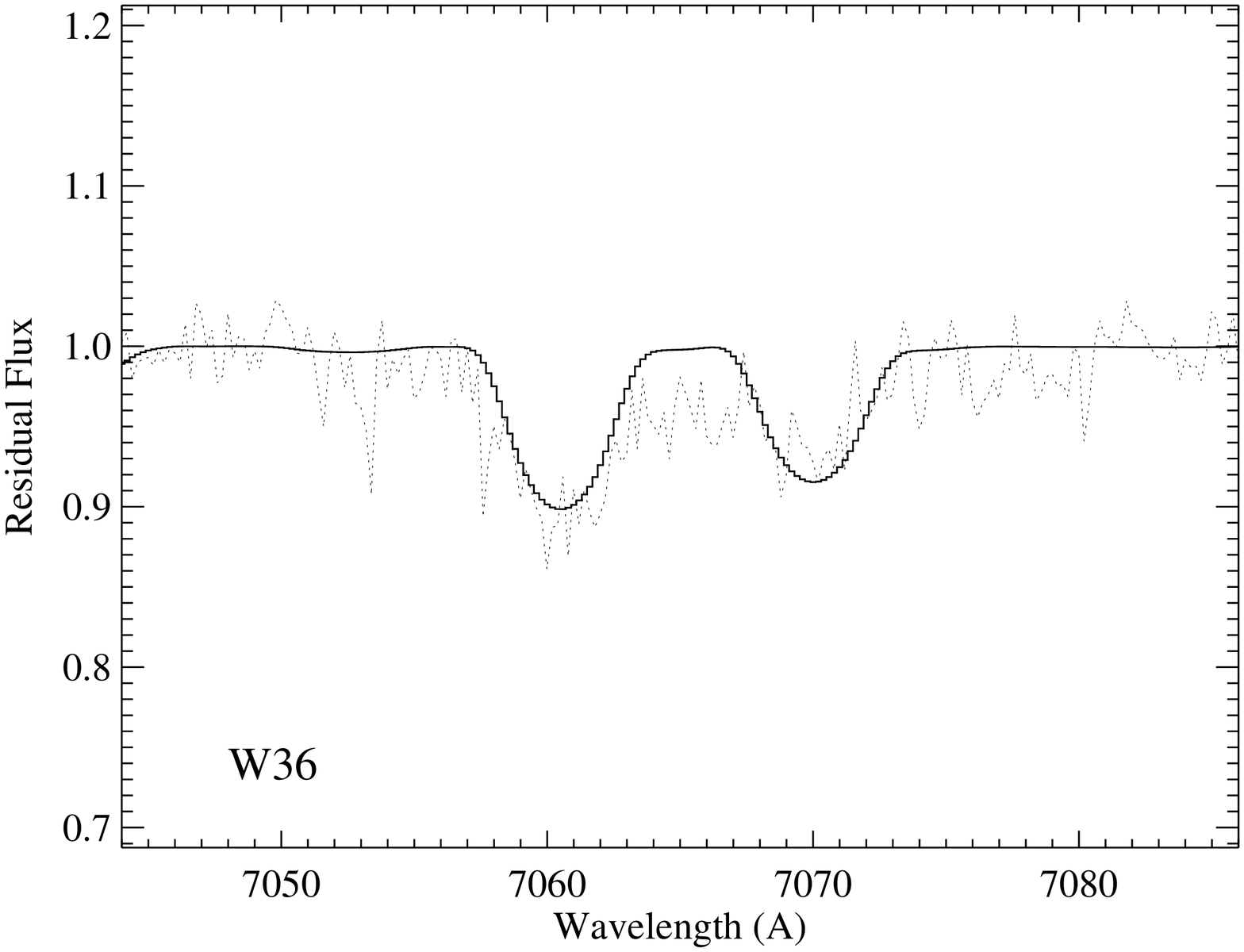} \\ \includegraphics[width=2.5in, angle=90]{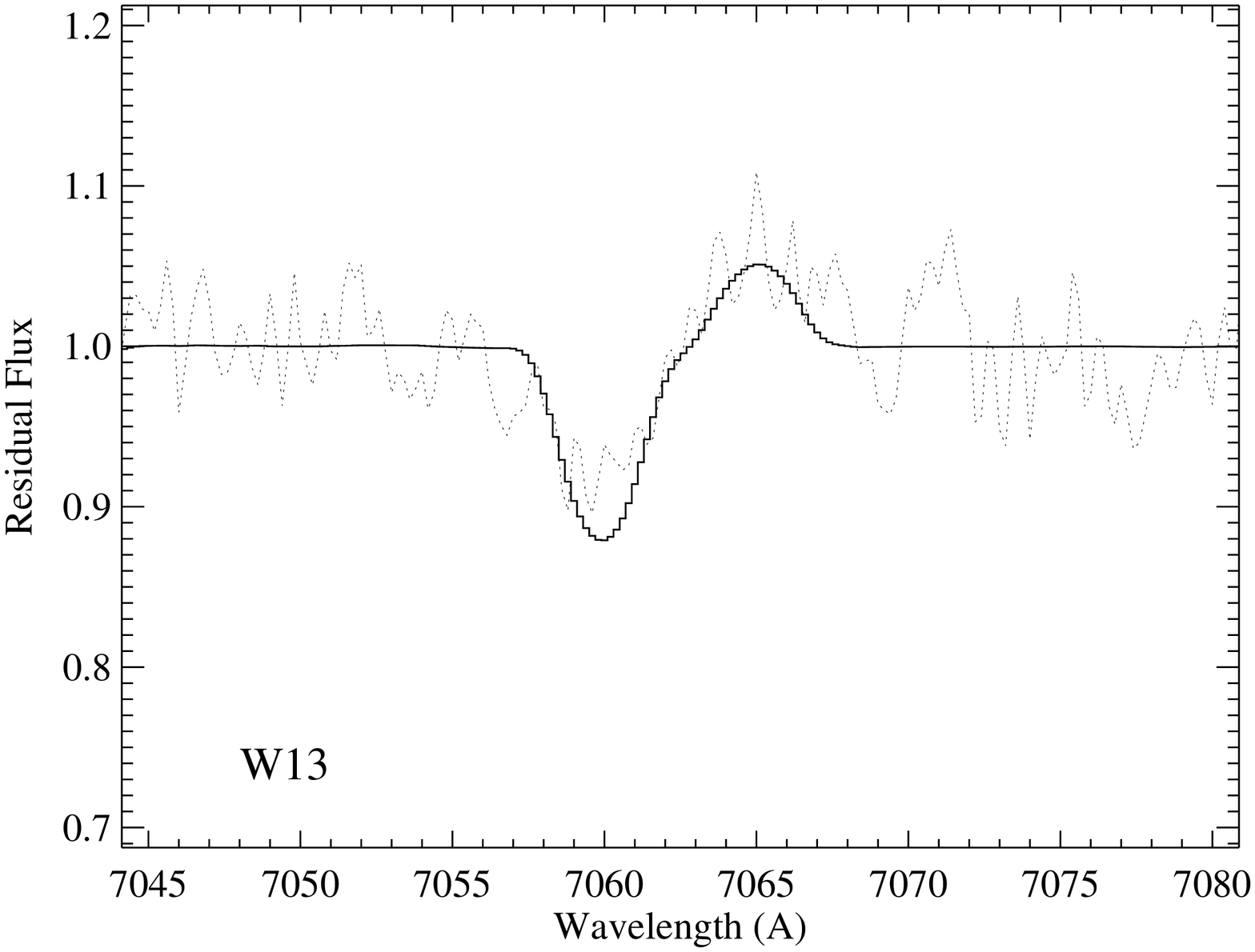} &
\includegraphics[width=2.5in, angle=90]{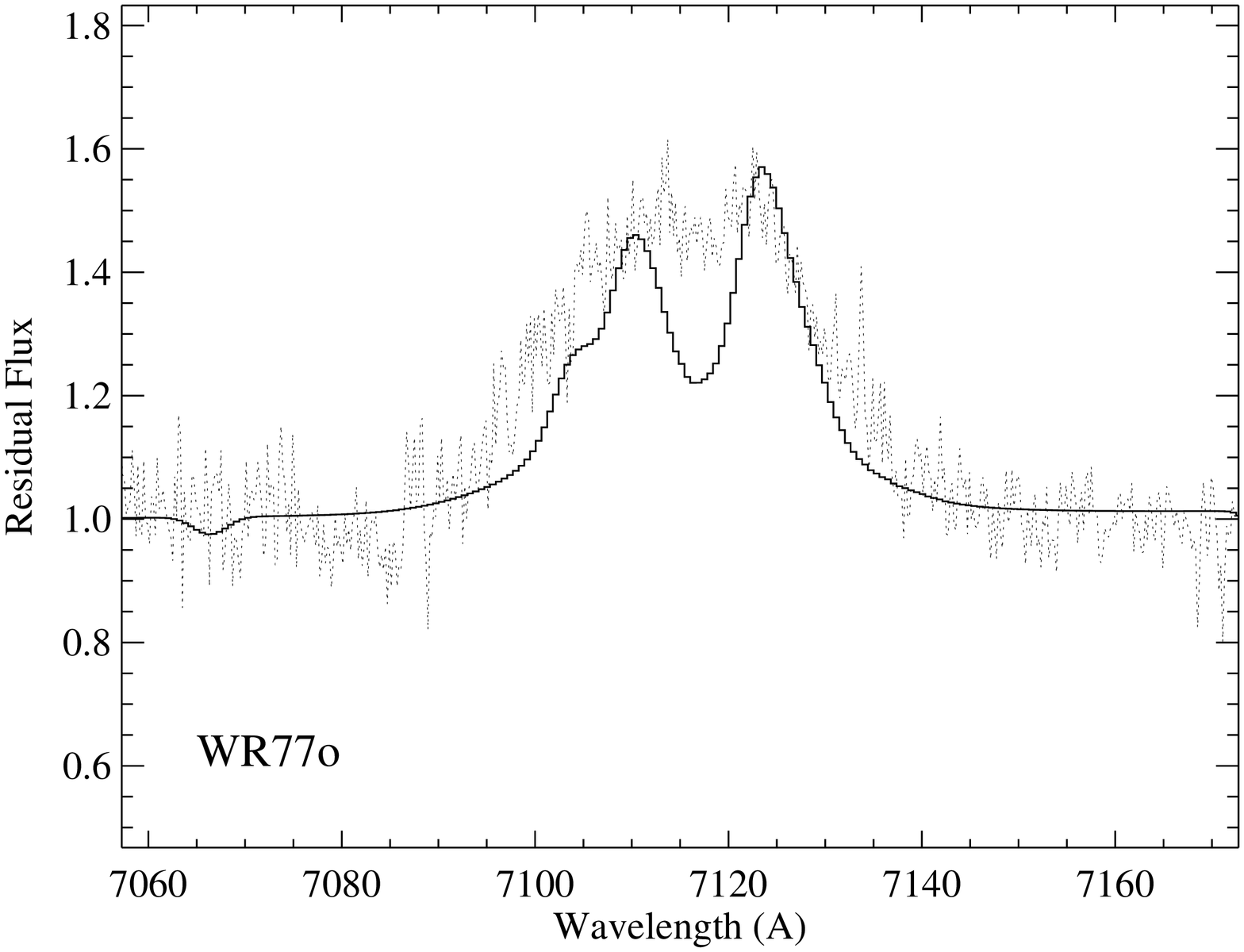}
\end{array}$
\end{center}
\caption{Representative examples of the model fit (solid line) to the observed spectra (dotted lines), showing HeI 7065 for W13, W$_{DEB}$, W36 and NIV 7103--7128 for WR77o.}
\label{fig:fit}
\end{figure*}

The radial velocities (RVs) were determined with a $\chi^2$ minimization
method, which compares the observed spectrum to a grid of synthetic
spectra. 
\begin{table}[h]
\caption{Radial velocity measurements}

\centering
\small\addtolength{\tabcolsep}{-4.5pt}
\begin{tabular}{c c c c c c c }
\hline\hline
Binary & HJD & Phase & RV$_{1}$ $\pm \,\sigma_{RV_{1}}$ & $(O-C)_1$ & RV$_{2}$ $\pm \,\sigma_{RV_{2}}$ & $(O-C)_2$ \\
 & $-$2450000 & $\phi$ & (km~s$^{-1}$) & (km~s$^{-1}$) & (km~s$^{-1}$) &
(km~s$^{-1}$) \\
\hline\hline

W$_{DEB}$ & 4167.87328 & 0.42 & $-$188$\pm$35 & $-$19 & 105$\pm$11 & $-$ 16 \\
 & 4168.86327 & 0.64 & 95$\pm$35 & 56 & $-$165$\pm$11  & $-$17 \\
 & 4620.60505 & 0.22 & $-$205$\pm$35 & $-$40  & 63$\pm$11  & $-$54 \\
 & 5805.52158 & 0.66 & 130$\pm$35 & 75 & $-$165$\pm$11 & $-$6 \\ \hline

W36 & 4160.79354 & 0.90 &    100$\pm$32  & 27  &$-$200$\pm$16  & $-$16  \\
 & 4167.79030 & 0.10 & $-$170$\pm$32  & $-$25  &   110$\pm$16  & 4   \\
 & 4168.78891 & 0.41 & $-$160$\pm$32  & $-$33 &    95$\pm$16  & $-$10  \\
 & 4174.78780 & 0.30  & $-$200$\pm$32  &$-$3 & 170$\pm$16  & $-$24  \\
 & 4280.76847 & 0.61 &   100$\pm$32  & 25  &$-$200$\pm$16  & $1$  \\
 & 4282.66772 & 0.21 & $-$180$\pm$32  & 19 & 165$\pm$16  & $-$34  \\
 & 4708.56044 & 0.09 &   \ldots  & \ldots  &          100$\pm$16  & 0  \\
 & 4710.55897 & 0.72 &   100$\pm$32  & $-$27  &$-$270$\pm$16  & 5   \\ \hline

W13 & 4160.77158 & 0.88 &  \ldots  & \ldots    &  $-$155$\pm$13    & 7  \\
 & 4167.76394 & 0.64 &   85$\pm$23    & $-$16  &  $-$190$\pm$13 & $-$20  \\
 & 4168.76923 & 0.74 &  145$\pm$23    & $-$2  &  $-$210$\pm$13 & $-$1  \\ 
 & 4174.76249 & 0.39 & $-$180$\pm$23  & $-$7  &  13$\pm$13     & $-$7  \\
 & 4279.61631 & 0.71 &  \ldots  &    \ldots       &  $-$230$\pm$13 & $-$28  \\
 & 4280.58040 & 0.81 &  \ldots  &    \ldots       &$-$180$\pm$13   & 17  \\
 & 4281.62310 & 0.92 &  \ldots  &    \ldots        &$-$125$\pm$13   & 6  \\
 & 4708.48711 & 0.99 &  \ldots  &  \ldots         &$-$95$\pm$13    & $-$19  \\
 & 4710.48815 & 0.20 & $-$250$\pm$23 & $-$11  &   75$\pm$13    & 8  \\ \hline

WR77o &4160.84107 & 0.98 & \ldots & \ldots & $-$25 $\pm$30  &     58    \\
 &4167.83559 & 0.97 & \ldots & \ldots & $-$15 $\pm$30  &      97   \\
 &4168.82036 & 0.25 & \ldots & \ldots &  290 $\pm$30  &     $-$35    \\
 & 4174.82619 & 0.95 & \ldots & \ldots & $-$35 $\pm$30  &     108    \\
 &4282.49396 & 0.54 & \ldots & \ldots & $-$230 $\pm$30  &   $-$100      \\
 &4620.66407 & 0.61 & \ldots & \ldots & $-$340 $\pm$30  &    $-$74     \\
 &4708.52114 & 0.57 & \ldots & \ldots & $-$250 $\pm$30  &     $-$61    \\
 &4710.51970 & 0.14 & \ldots & \ldots &  255 $\pm$30  &      13   \\
\hline

\end{tabular}
\label{rv}
\end{table}
Each synthetic spectrum was constructed as a sum of two models, each shifted
over a range of radial velocities with a velocity step of 5~km~s$^{-1}$. We used TLUSTY models from the
OSTAR2002 by \citet{Lanz03} and BSTAR2006 by \citet{Lanz07} grids for the 
components of each OB-type system. The
parameters of the synthetic spectra for each system were selected to
reproduce the observed spectra and the rotational velocities ($v\, \sin i$) were adopted to be 80~km~s$^{-1}$ and 110~km~s$^{-1}$ for
the cooler and the hotter star of W$_{DEB}$, respectively, and 110~km~s$^{-1}$ for the components of W36 and W13. We inverted the TLUSTY model in order to match the emission line component of W13. Whilst this approach might seem somewhat unusual, it turned out to provide more stable results than a two Gaussian fit. The $\chi^2$ was computed as the quadratic sum
of the differences between the observed and the synthetic spectrum 
for a range of radial velocities. We computed the minimization using
60$\AA$ centered on each of the two available narrow helium lines
($\lambda\lambda6678,7065$), which yield more accurate radial velocities 
compared to the broader hydrogen lines. For WR77o, we used a WNE (early type) 
PoWR model \citep{Graefener02,Hamann03,Hamann04}. Each PoWR model depends on two parameters: the stellar temperature T* and the transformed radius $R_{t}$ \citep[defined by][]{Schmutz92} which takes into account the stellar radius, the mass--loss rate and the terminal--wind velocity. We adopted a model with a stellar temperature of T* = $50\,100$~K and transformed radius $\log{R_{t}}=0.7$ and computed the minimization 
using 140 \,\AA\, centered on 
the 
\ion{N}{IV}~$\lambda\lambda7103-7128$ line. The resulting velocities and
errors for all four systems are given in Table~\ref{rv}. The fit producing the lowest $\chi^2$ yielded the best fit model of the grid, and therefore the radial velocities for each component. The errors were determined using the $\sigma$ values, which were the mean values of the difference between the data and the model radial velocity curves determined in the following section. Figure~\ref{fig:fit} shows representative examples of the model fit to the observed spectra for each binary system, illustrating the good quality of the fit.

\section{Binary modeling} 

With the radial velocity measurements at hand, we proceeded to fit the
$V,\,R$ and $I-$band light curves published by \citet{Bonanos07} and the radial
velocity curves using the PHOEBE interface \citep{Prsa05} to the
Wilson-Devinney code \citep[WD;][]{Wilson71}. For all the binaries, we used the
square root limb--darkening law, which works better for hot stars at
optical wavelengths. Bolometric and passband limb--darkening
coefficients were taken from the in--house computed limb--darkening PHOEBE 2011 tables.
We fixed albedos 
and gravity brightening exponents to unity from
theoretical values for stars with radiative
envelopes. 

The errors for each parameter were determined with the phoebe--scripter, 
following \citet{Bonanos09}. 
The script uses the WD differential corrections algorithm to perform the 
minimization 
for the input values, adopts the minimizer results by applying the corrections to the parameter table and
repeats these steps 1000 times. By calculating the $\chi^{2}$ 
minimization for each iteration we end up with statistics from which 
the errors for each parameter can be determined. 

Below we discuss each system separately and present the resulting
orbital and astrophysical parameters of the components. 

\begin{figure*}[h]
\begin{center}$  
\includegraphics[scale=0.3]{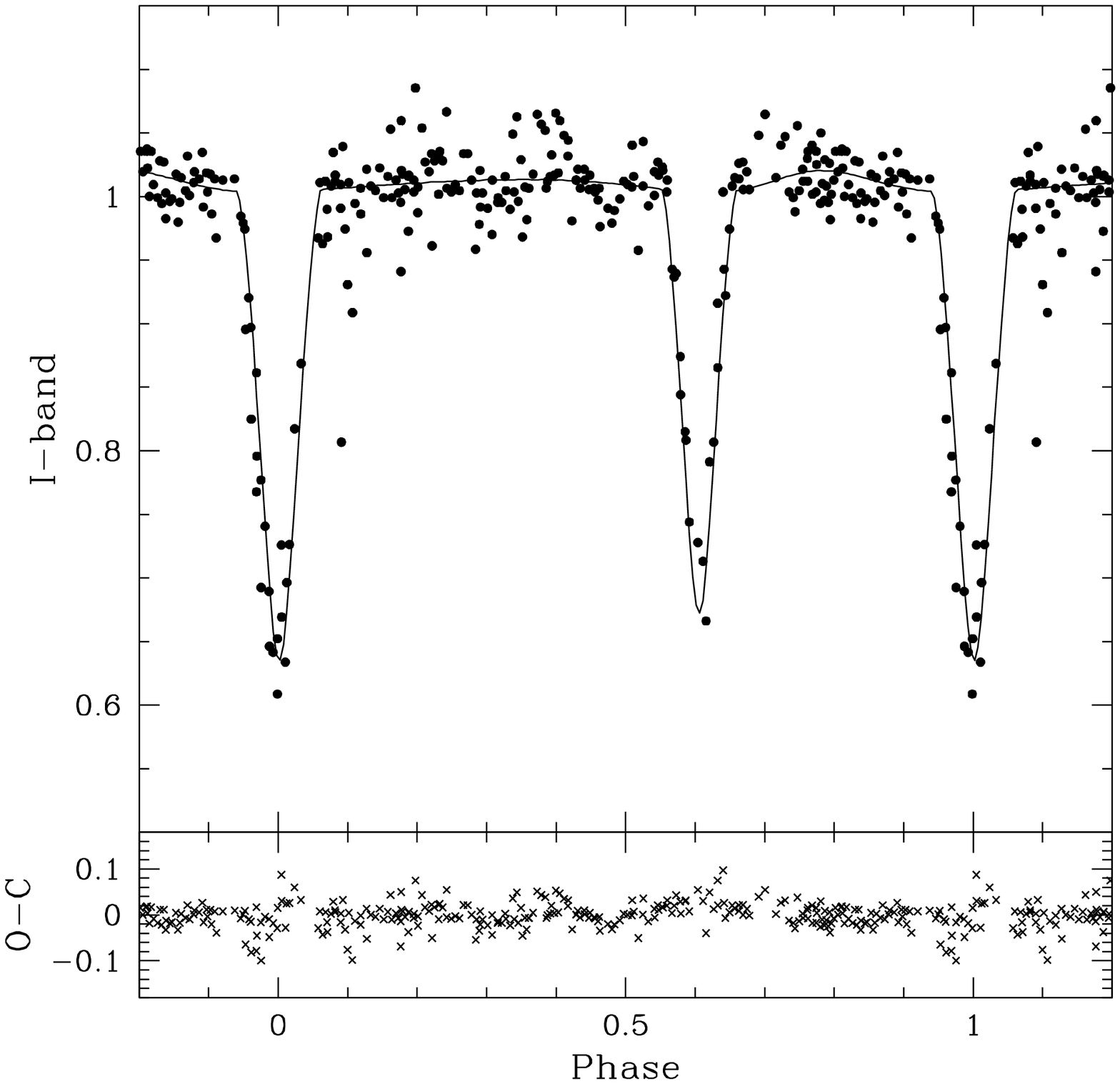}\includegraphics[scale=0.3]{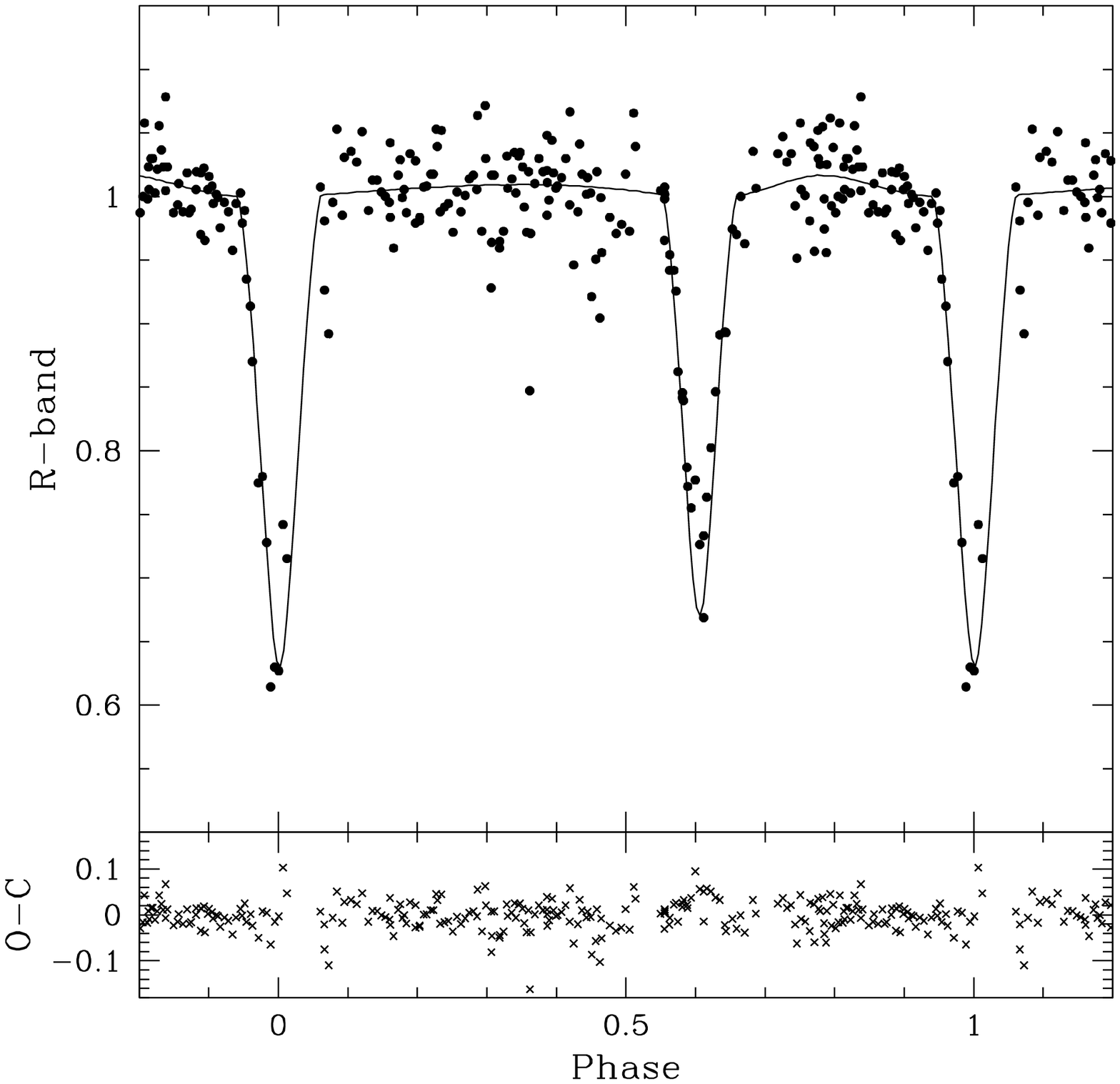}\includegraphics[scale=0.3]{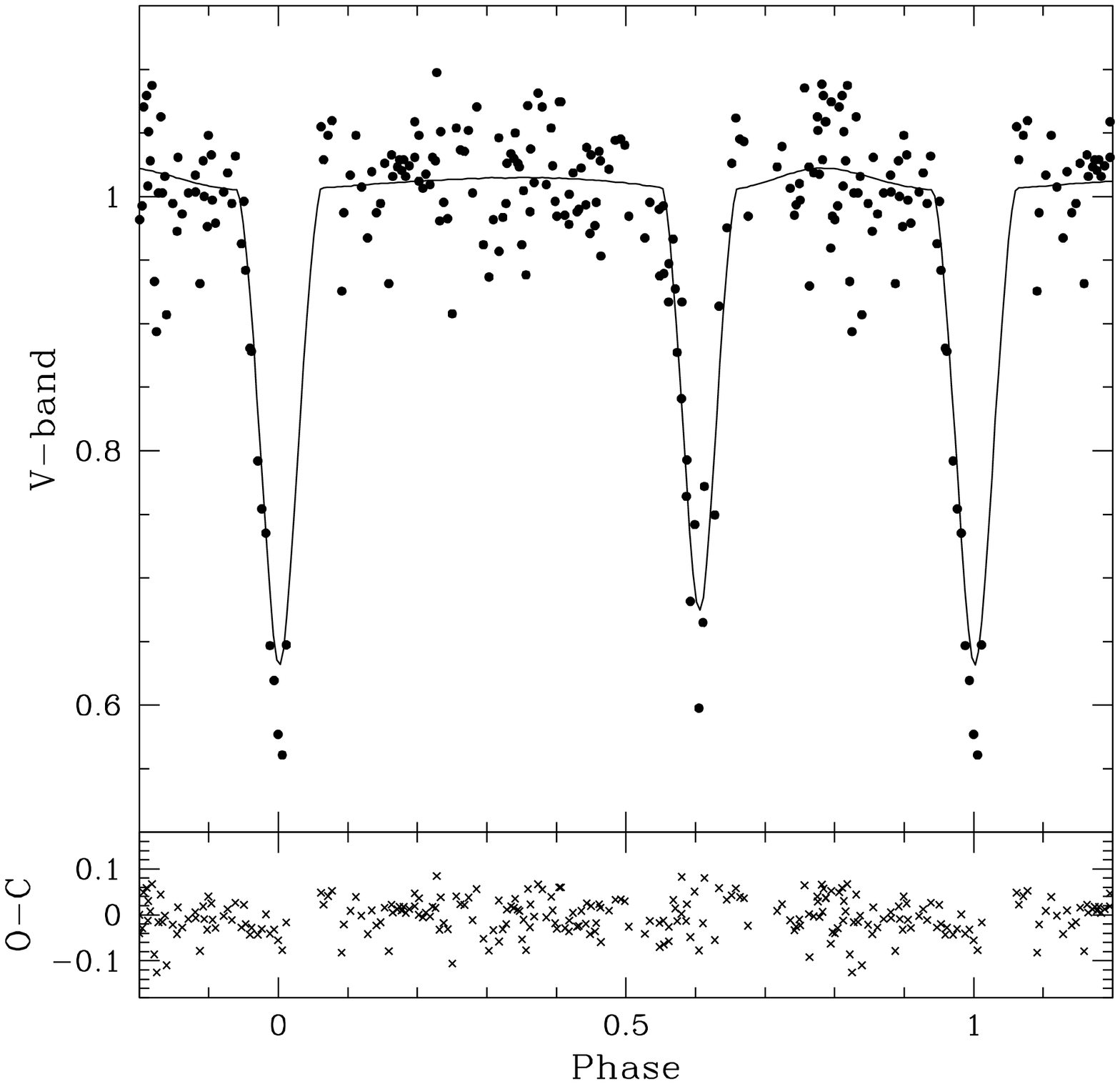}$
\end{center}
\caption{Phased light curves of W$_{DEB}$ in the $I, R$ and $V-$band in units of normalized light, with best fit PHOEBE model overplotted (detached configuration). The apparent magnitudes outside of eclipse are $I=14.8$, $R=17.35$ and $V=20.3$. The lower panel in each light curve shows the $O-C$ residuals.}
\label{fig:lcwdeb}
\end{figure*}

\subsection{W$_{DEB}$} 

The Westerlund~1 detached eclipsing binary
(W$_{DEB}$\footnote{RA=16:46:58.78, Dec=$-$45:54:31.9, J2000}) is a
double-lined, eccentric, detached system with a period of
4.4473 d. W$_{DEB}$ is located 4$\arcmin$ south of the cluster core, 
but its colors are similar to those of the other cluster
stars, a fact that indicates that this binary is a member of Wd~1. The
components of W$_{DEB}$ were the first main sequence stars to be identified
in Westerlund~1 \citep{Bonanos07} and as such, their parameters can 
be compared to models
of single star evolution and be used to estimate the age of the cluster.

We obtained the first spectra for this system, which reveal an OB
spectral type and confirm cluster membership. However, given the wavelength range available, a more accurate determination of their
exact spectral types and effective temperatures was not
possible. 
We therefore adopted a temperature of $25\,000K$ in the
analysis below as a starting point for our fitting. The few useful spectra (four) 
that we could work with, provide us with parameters that
suffer from large errors ($>10\%$).

\begin{table}[h]
\caption{Results for W$_{DEB}$ from combined $LC$ and $RV$ curve analysis}

\centering
\small\addtolength{\tabcolsep}{-3.0pt}
\begin{tabular}{l c }
\hline\hline
Parameter & Value \\
\hline\hline
Period, $P$ & 4.4473 $\pm$ 0.0004 $d$ \\
Time of primary eclipse, \emph{HJD$_0$} & 2453934.741 $\pm$ 0.003  \\
Inclination, $i$        & 86 $\pm$ 2 $\deg$\\
Eccentricity, $e$ & 0.175 $\pm$ 0.003 \\
Longitude of periastron, $\omega$ & 252 $\pm$ 1 $deg$ \\
Surface potential, $\Omega_{1}$ & 5.9 $\pm$ 0.1  \\
Surface potential, $\Omega_{2}$ & 6.50 $\pm$ 0.06 \\
Light ratio in $V$, $L_{2}/L_{1}$ & 0.499 $\pm$  0.005 \\
Light ratio in $R$, $L_{2}/L_{1}$ & 0.501 $\pm$  0.005 \\
Light ratio in $I$, $L_{2}/L_{1}$ & 0.505 $\pm$ 0.005 \\
Mass ratio, $q$ & 0.80 $\pm$ 0.07   \\
Systemic velocity, $\gamma$ & $-40\pm3\; \rm km\; s^{-1}$ \\
Semi-major axis, $a$ & $34 \pm1\; \emph{\rsun} $  \\
Semi-amplitude, $K_{1}$ & $229\pm19\; \rm km\; s^{-1}$ \\
Semi-amplitude, $K_{2}$ & $187\pm6\; \rm km\; s^{-1}$ \\
Radius$^{\dagger}$, $\rm r_{1,pole}$ & 0.202   $\pm$ 0.006 \\     
............ $\rm r_{1,point}$& 0.208 $\pm$ 0.007    \\
............ $\rm r_{1,side}$ & 0.203  $\pm$ 0.006   \\
............ $\rm r_{1,back}$ & 0.206  $\pm$ 0.007   \\
............ $\rm r_{1}$$^{*}$ & 0.204 $\pm$ 0.006   \\
Radius$^{\dagger}$, $\rm r_{2,pole}$ & 0.154   $\pm$ 0.005  \\     
............ $\rm r_{2,point}$& 0.157   $\pm$ 0.006  \\
............ $\rm r_{2,side}$ & 0.155   $\pm$ 0.006  \\
............ $\rm r_{2,back}$ & 0.157   $\pm$ 0.006  \\
............ $\rm r_{2}$$^{*}$ & 0.156 $\pm$ 0.006 \\
\hline
Notes: \\
$\dagger$ The radii are given in units of \emph{a}. \\ $*$ Mean radius. 
\end{tabular}
\label{tab:wddeb}
\end{table}

We ran PHOEBE, using the three light curves and the radial velocity curves, in the detached mode for the following free parameters:
the inclination \emph{i}, eccentricity \emph{e}, the argument of periastron
\emph{$\omega$}, the semi-major axis \emph{$a$}, the radial velocity of the center
of mass \emph{$\gamma$}, the surface potential for each star \emph{$\Omega$}, 
the mass ratio \emph{q}=M$_{2}$/M$_{1}$,
the effective temperature \emph{$T_\mathrm{eff}$}, the period \emph{P}, 
the time of primary eclipse \emph{HJD$_{0}$} and
the band pass luminosity \emph{$L_{2}/L_{1}$} for each band. We first fit the light curve 
to estimate \emph{P}, \emph{HJD$_{0}$}, \emph{e}, \emph{$\omega_{0}$}, 
\emph{i}, 
\emph{$\Omega$} for each star, then fixed these values and fit the radial velocity curve 
separately to estimate \emph{q}, \emph{$a$} and \emph{$\gamma$}. 

\begin{figure}[h]
\includegraphics[scale=0.33]{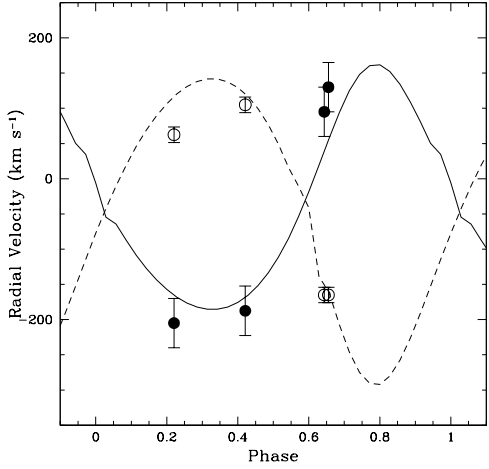}
\caption{Radial velocity curve for W$_{DEB}$. The measurements are shown as filled circles for the primary and open circles for the secondary; overplotted is the best-fit model from PHOEBE, denoted by a solid line for the primary and a dashed line for the secondary.}
\label{fig:rvwdeb}
\end{figure}

For a detached, eccentric system with partial eclipses such as W$_{DEB}$,
the ratio of radii cannot be constrained only by photometry, so we used 
the spectroscopic light ratio to determine the primary to be the larger 
and hotter star. Following \citet{North10}, we next ran 
PHOEBE keeping the surface potential for the secondary component constant and then similarly determined the 
surface potential for the primary component, while we adopted the value for \emph{q} that was determined from the radial 
velocity curve. After the initial fit to the light curves and the radial velocity curve, we iteratively refit the light curves and subsequently the radial velocity curve, until the values converged. The last step was to run PHOEBE for all the
free parameters described above, but keeping \emph{q} and \emph{$\Omega_{1}$} fixed. Table~\ref{tab:wddeb}
presents the parameters resulting from the analysis with PHOEBE. The light and radial velocity
curve fits are shown in  Figure~\ref{fig:lcwdeb} and~\ref{fig:rvwdeb}, respectively.  
The primary star, corresponding to the hotter component, was found to have a mass of $15\,\emph{\msun}$ and radius of $7\,\emph{\rsun}$ 
while the secondary was found to
have $12\,\emph{\msun}$ and \clearpage $5\,\emph{\rsun}$. The $\log{g}$ 
values of the two stars were found to be $\sim 4$ which confirms the main sequence nature of the components.
In addition, the systemic velocity of \emph{$\gamma$} = $-$40~km~s$^{-1}$, which is similar to that measured for the other three systems, provides further evidence for the membership of W$_{DEB}$ to
the cluster. Table~\ref{tab:bv} presents the physical parameters (mass, radius, $\log{g}$, $T_\mathrm{eff}$ and $\log(\rm L/\emph{\lsun})$) derived for each eclipsing binary. 

\subsection{W36} 

W36\footnote{RA=16:47:05.08, Dec=$-$45:50:55.1, J2000} is the second
brightest of our targets, with no previous spectroscopy available. It is
a contact double-lined eclipsing binary in a circular orbit with a
period of 3.1811 d. The spectra of this system indicate that the
components are of spectral type OB. As was the case for W$_{DEB}$, the determination of their exact
spectral types and thus of their accurate temperatures was not possible.

\begin{figure}[h]  
\includegraphics[scale=0.33]{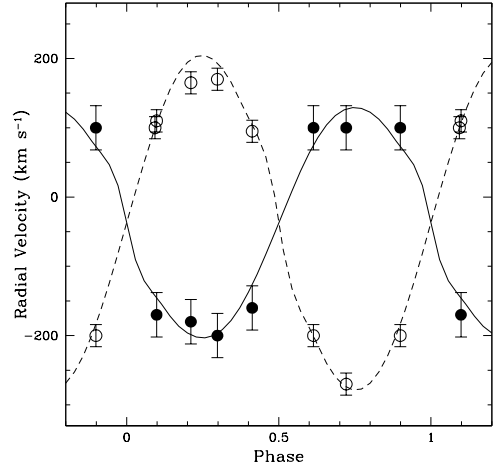}
\caption{Radial velocity curve for W36. The measurements are shown as filled circles for the primary and open circles for the secondary; overplotted is the best-fit model from PHOEBE, denoted by a solid line for the primary and a dashed line for the secondary.}
\label{fig:rvw36}
\end{figure}

\begin{figure}[h]
\begin{center}$
\begin{array}{c}  
\includegraphics[scale=0.3]{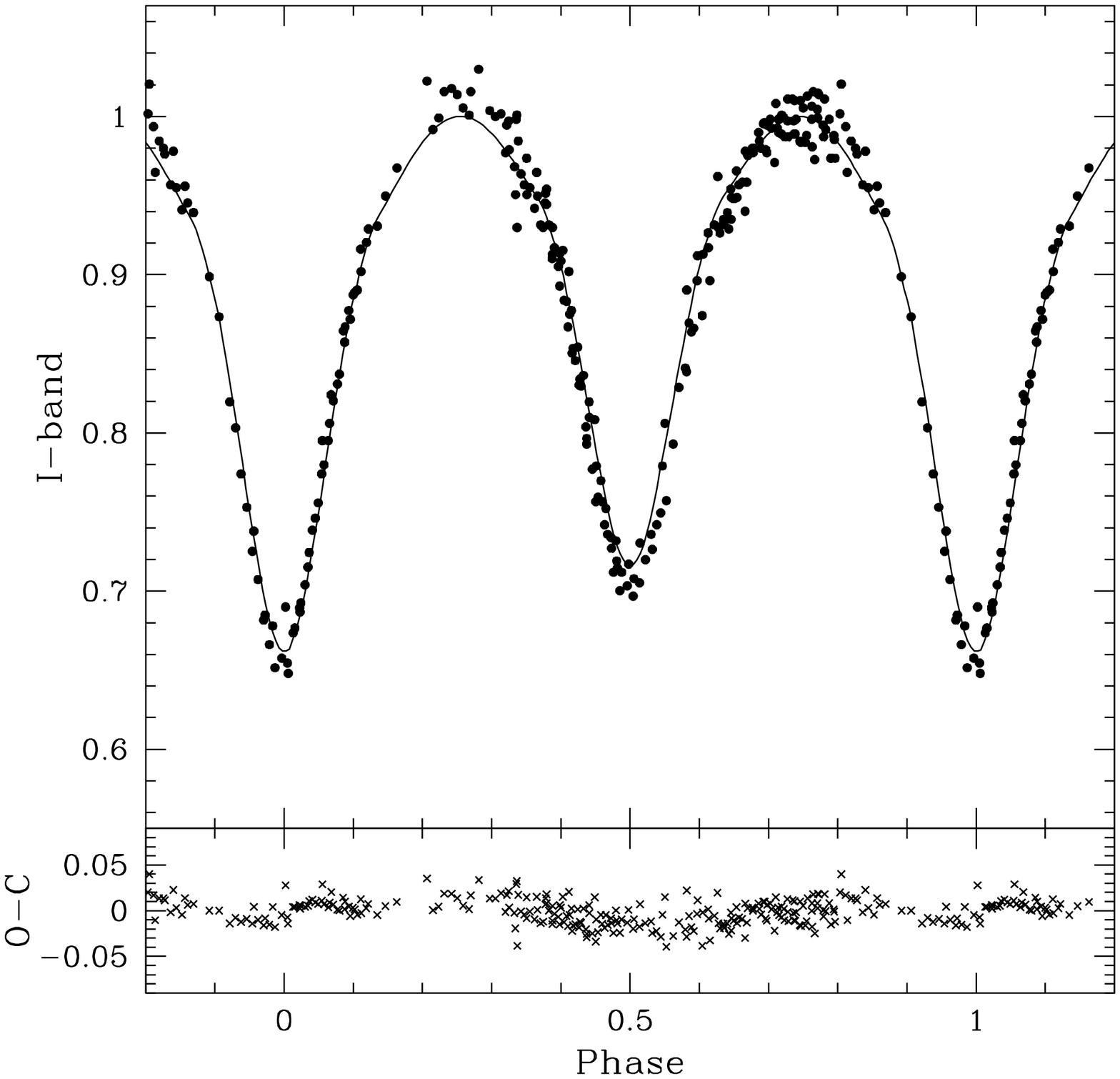} \\ \includegraphics[scale=0.3]{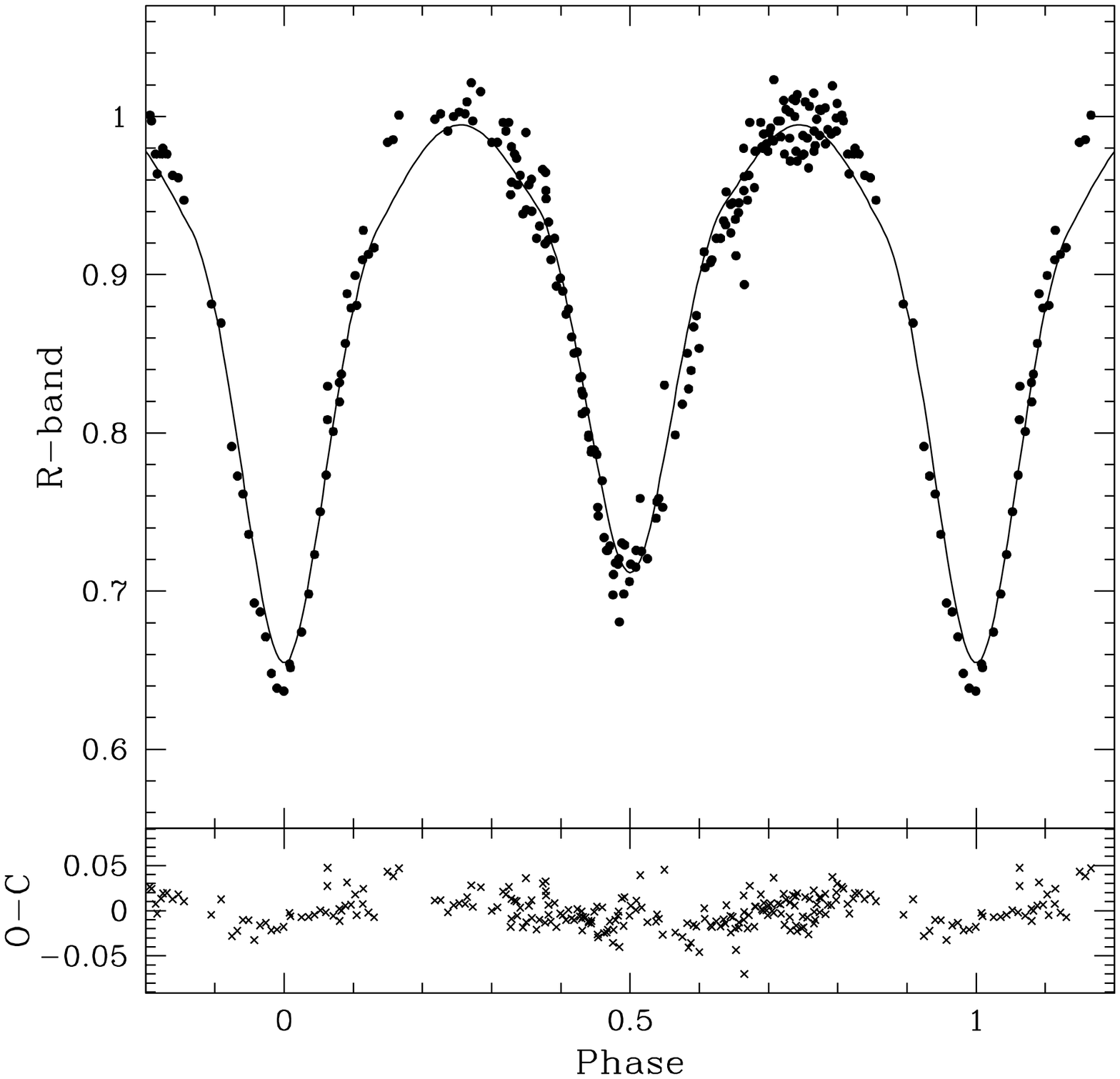} \\ \includegraphics[scale=0.3]{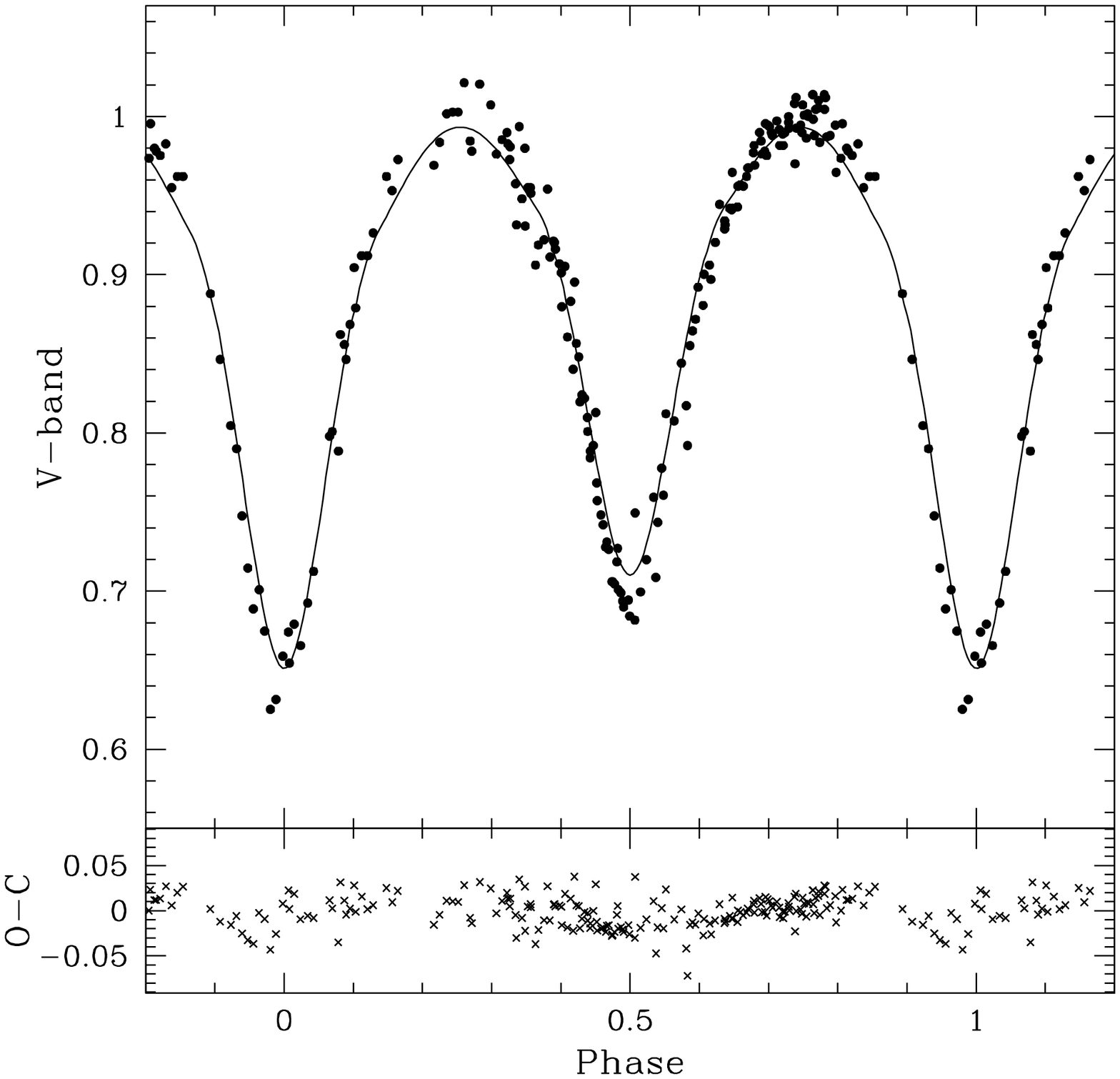}
\end{array}$
\end{center}
\caption{Phased light curves of W36 in the $I, R$ and $V-$band in units of normalized light, with best fit PHOEBE model overplotted (contact configuration). The apparent magnitudes outside of eclipse are $I=13.1$, $R=15.8$ and $V=18.85$. The lower panel in each light curve shows the $O-C$ residuals.}
\label{fig:lcw36}
\end{figure}

We ran PHOEBE, using the three light curves and the radial velocity curves, 
in the contact mode for the following free
parameters: the inclination \emph{i}, the semi-major axis \emph{$a$}, the radial
velocity of the center of mass \emph{$\gamma$}, the surface potential for the
primary star \emph{$\Omega_{1}$}, the mass ratio \emph{q}, the effective temperature 
of each star \emph{$T_\mathrm{eff}$},
the period \emph{P}, the time of primary eclipse \emph{HJD$_{0}$} and the band pass luminosity
\emph{$L_{2}$/$L_{1}$} for each band. Table~\ref{tab:w36} presents the parameters
resulting from the analysis with PHOEBE. The radial velocity and light
curves are shown in Figure~\ref{fig:rvw36} and~\ref{fig:lcw36}, respectively.

The primary star of this system was found to have a mass of 16 $\,\emph{\msun}$ 
and radius of 11 $\,\emph{\rsun}$, while the secondary has 11 $\,\emph{\msun}$
and 9 $\,\emph{\rsun}$ (see Table~\ref{tab:bv}). Since it is a contact system the 
masses that were determined are not the initial masses, as mass transfer has 
occurred. Values of $\log{g}<4$ found for both stars also testify to the evolved nature 
of the two stars. 

\begin{table}[h]
\caption{Results for W36 from combined $LC$ and $RV$ curve analysis}

\centering
\small\addtolength{\tabcolsep}{-3.0pt}
\begin{tabular}{l c }
\hline\hline
Parameter & Value \\ 
\hline\hline
Period, $P$ & 3.18110 $\pm$ 0.00003 $d$ \\
Time of primary eclipse, \emph{HJD$_0$} & 2453909.8080 $\pm$ 0.0005 \\
Inclination, $i$ & 73.0 $\pm$ 1.7 $\deg$\\
Surface potential, $\Omega_{1}$ & 3.29 $\pm$ 0.03  \\
Surface potential, $\Omega_{2}$ & 3.28 (fixed) \\
Light ratio in $I$, $L_{2}/L_{1}$ & 0.514 $\pm$  0.005 \\
Light ratio in $R$, $L_{2}/L_{1}$ & 0.507 $\pm$  0.005 \\
Light ratio in $V$, $L_{2}/L_{1}$ & 0.504 $\pm$ 0.005 \\
Mass ratio, $q$ & 0.69 $\pm$ 0.03 \\
Systemic velocity, $\gamma$ & $-37\pm2\; \rm km\; s^{-1}$ \\
Semi-major axis, $a$ & $27.5 \pm1.5\; \emph{\rsun} $  \\
Semi-amplitude, $K_{1}$ & $175\pm16\; \rm km\; s^{-1}$ \\
Semi-amplitude, $K_{2}$ & $235\pm8\; \rm km\; s^{-1}$ \\
Radius$^{\dagger}$, $\rm r_{1,pole}$ & 0.378   $\pm$ 0.003 \\     
............ $\rm r_{1,point}$& 0.47   $\pm$ 0.01  \\
............ $\rm r_{1,side}$ & 0.398   $\pm$ 0.003  \\
............ $\rm r_{1,back}$ & 0.423   $\pm$ 0.004  \\
............ $\rm r_{1}$$^{*}$ &  0.400 $\pm$ 0.004 \\
Radius$^{\dagger}$, $\rm r_{2,pole}$ & 0.316   $\pm$ 0.003  \\     
............ $\rm r_{2,point}$& 0.40   $\pm$ 0.01  \\
............ $\rm r_{2,side}$ & 0.330   $\pm$ 0.004  \\
............ $\rm r_{2,back}$ & 0.357   $\pm$ 0.005  \\
............ $\rm r_{2}$$^{*}$ & 0.334 $\pm$ 0.005\\
\hline
Notes: \\
$\dagger$ The radii are given in units of \emph{a}. \\ $*$ Mean radius. 
\end{tabular}
\label{tab:w36}
\end{table}

\vspace{0.9cm}

\subsection{W13} 

W13\footnote{RA=16:47:06.46, Dec=$-$45:50:26.0, J2000} is our brightest
target: a near--contact double--lined eclipsing binary in a circular orbit
with a period of 9.266 d.  The spectra indicate an emission plus absorption line
object, consistent with the spectral types B0.5Ia+/WNVL and O9.5-B0.5I, respectively, reported by the independent study of the system by \citet{Ritchie10}. The known spectral types provide
additional constraints to determine the
fundamental parameters of each component. In addition, the well determined temperatures of
the members of W13 make this system the best available eclipsing binary in Wd~1 for a direct distance determination.
 
We ran PHOEBE, using only the less noisy $R$-band light curve and the radial velocity curves, in the detached mode, as the components are not filling
the Roche lobes, for the following free parameters: \emph{i}, \emph{$a$}, \emph{$\gamma$}, 
\emph{$\Omega$}, \emph{q}, \emph{P}, \emph{HJD$_{0}$}, \emph{$T_\mathrm{eff_2}$} and \emph{$L_{1}$}. 
Our radial velocity curves contain both our own RV measurements plus those made by \citet{Ritchie10}. Table~\ref{tab:w13} presents the parameters resulting
from the analysis with PHOEBE. 

\begin{table}[h]
\caption{Results for W13 from combined $LC$ and $RV$ curve analysis}

\centering
\small\addtolength{\tabcolsep}{-3.0pt}
\begin{tabular}{l c }
\hline\hline
Parameter & Value \\
\hline\hline
Period, $P$ & 9.2665 $\pm$ 0.0003 $d$ \\
Time of primary eclipse, \emph{HJD$_0$} & 2453902.3998 $\pm$ 0.0002 \\
Inclination, $i$  & 61 $\pm$ 2 $\deg$\\
Surface potential, $\Omega_{1}$ & 4.6 $\pm$ 0.2  \\
Surface potential, $\Omega_{2}$ & 5.6 $\pm$ 0.2 \\
Light ratio in $R$, $L_{2}/L_{1}$ & 0.822 $\pm$ 0.005 \\
Mass ratio, $q$ & 1.42 $\pm$ 0.04 \\
Systemic velocity, $\gamma_{emis}$ & $-48.2\; \rm km\; s^{-1}$ (fixed) \\
Systemic velocity, $\gamma_{abs}$ & $-65.9\; \rm km\; s^{-1}$ (fixed) \\
Semi-major axis, $a$ & $71 \pm1\; \emph{\rsun} $  \\
Semi-amplitude, $K_{1}$ & $198\pm7\; \rm km\; s^{-1}$ \\
Semi-amplitude, $K_{2}$ & $142\pm4\; \rm km\; s^{-1}$ \\
Radius$^{\dagger}$, $\rm r_{1,pole}$ & 0.31   $\pm$ 0.01 \\     
............ $\rm r_{1,point}$& 0.37   $\pm$ 0.04  \\
............ $\rm r_{1,side}$ & 0.32   $\pm$ 0.02  \\
............ $\rm r_{1,back}$ & 0.34   $\pm$ 0.02  \\
............ $\rm r_{1}$$^{*}$ &  0.32 $\pm$ 0.02 \\
Radius$^{\dagger}$, $\rm r_{2,pole}$ & 0.29   $\pm$ 0.03  \\     
............ $\rm r_{2,point}$& 0.31   $\pm$ 0.03  \\
............ $\rm r_{2,side}$ & 0.30   $\pm$ 0.03  \\
............ $\rm r_{2,back}$ & 0.31   $\pm$ 0.03  \\
............ $\rm r_{2}$$^{*}$ & 0.30 $\pm$ 0.03 \\
\hline
Notes: \\
$\dagger$ The radii are given in units of \emph{a}. \\ $*$ Mean radius. 
\end{tabular}
\label{tab:w13}
\end{table}

\begin{figure}[h]  
\includegraphics[scale=0.33]{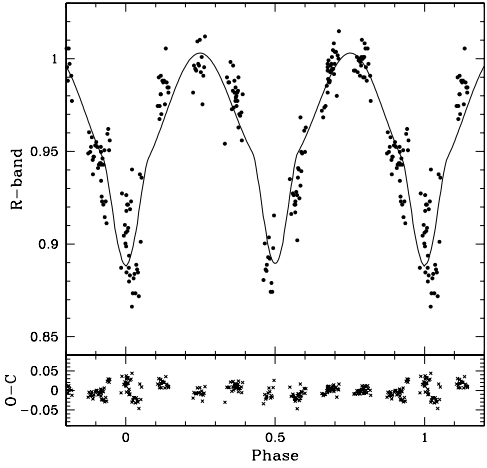}
\caption{Phased $R-$band light curve of W13 in units of normalized light, with best fit PHOEBE model overplotted (detached configuration). The apparent magnitude outside of eclipse is $R=14.65$. The lower panel in the light curve shows the $O-C$ residuals.}
\label{fig:lcw13}
\end{figure}

\begin{figure}[h]  
\includegraphics[scale=0.33]{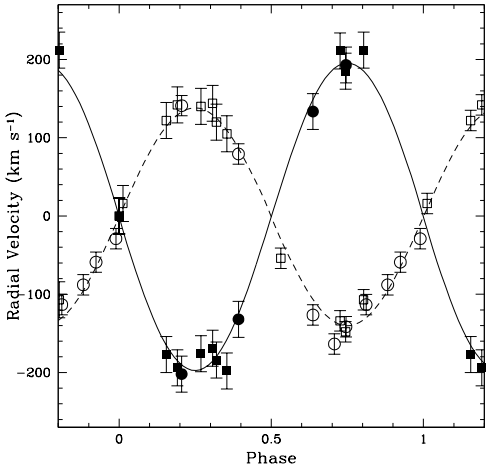}
\caption{Radial velocity curve for W13. The measurements are shown as filled circles for the primary and open circles for the secondary; overplotted is the best-fit model from PHOEBE, denoted by a solid line for the primary and a dashed line for the secondary. The measurements from \citet{Ritchie10} for the primary and secondary are denoted by filled and open squares, respectively. All radial velocities have been shifted to a systemic velocity of zero.}
\label{fig:rvw13}
\end{figure}

The values for the effective temperatures were taken from \citet{Ritchie10} and correspond to the spectral types they determined. More specifically, \emph{$T_\mathrm{eff_1}$} was fixed to 25\,000~K and we let \emph{$T_\mathrm{eff_2}$} be a free parameter with a starting value of 25\,000~K. The light and radial velocity
curves are shown in Figure~\ref{fig:lcw13} and~\ref{fig:rvw13}, respectively.

Moreover, for our analysis we adopted the $\gamma$ velocities for each 
component ($\gamma$$_{em}$ = $-$48~km~s$^{-1}$, $\gamma$$_{abs}$ = $-$66~km~s$^{-1}$) as found by \citet{Ritchie10}, and we transferred the 
whole system to 0 velocity by adjusting the measured velocities according to
the values above, since PHOEBE does not provide this option. In massive 
binaries, stellar winds can cause the measured $\gamma$ to differ for each star 
\citep{Massey77}. This fact is highly correlated to outward atmospheric mass 
motion and given that the components are supergiants, it was important to include this effect. 

The component that is responsible for the emission in our spectra 
was determined to have 23 $\pm$ 1 $\,\emph{\msun}$, 23 $\pm$ 2 $\,\emph{\rsun}$ and a filling factor $F=0.96$, values that are in agreement, although more precise than the values independently determined by \citet{Ritchie10}: 23.2 $^{+3.3}_{-3.0}$$\,\emph{\msun}$, 22 $\pm$ 2 $\,\emph{\rsun}$ and $F=0.93$. The component that causes the absorption 
in our spectra was determined to have 33 $\pm$ 2 $\emph{\msun}$, 21 $\pm$ 2 $\emph{\rsun}$ and $F=0.68$, compared to the values found by \citet{Ritchie10}: 35.4 $^{+5.0}_{-4.6}$$\,\emph{\msun}$, 21 $\pm$ 2 $\,\emph{\rsun}$ and $F=0.74$. The high values of filling factors of the two components, make W13 a near--contact system. It is worth mentioning that our RV measurements alone yielded results very close to those above, with the star that causes the emission having a mass of 23 $\pm$ 2 $\,\emph{\msun}$ and radius of 22 $\pm$ 2 $\,\emph{\rsun}$ and the one that causes the absorption having 32 $\pm$ 2 $\,\emph{\msun}$ and 23 $\pm$ 2 $\,\emph{\rsun}$. 
Furthermore we determined an inclination \emph{i} = 61 $\pm$ 2\degr \,, which is in good agreement with
the value 62 $\pm$ 4\degr \, that was determined from the binned light curve by \citet{Ritchie10}. The fact that we have almost doubled the number of RV measurements is the main reason that we determined a more accurate, but lower mass ratio \emph{q} = 1.42 $\pm$ 0.04 comparing to that of 1.53 $\pm$ 0.1 determined by \citet{Ritchie10}. Finally,
we improved the accuracy in the determination of the period to \emph{P} = 9.2665 $\pm$ 0.0003 d compared to $P$ = 9.2709 $\pm$ 0.0015 d determined by \citet{Ritchie10}.

\vspace{1.3cm}

\subsection{WR77o}

WR77o\footnote{RA=16:47:05.37, Dec=$-$45:51:04.7, J2000} is a
single--lined eclipsing binary in a circular orbit with a period of 3.52
d. The star that is visible in our spectra, is a
Wolf--Rayet star, reported to have a spectral type WN6-7
\citep{Negueruela05}, and later refined to WN7o \citep{Crowther06}.

For a single-lined spectroscopic binary, one can only determine the mass function of the system:

\begin{equation}
f(M_{1},M_{2}) = \frac{K^{3}_{2} P}{2 \pi G \sin^{3}{i}} = \frac{M^{3}_{1}}{(M_{1}+M_{2})^{2}}
\end{equation}

\noindent where \emph{K$_{2}$} is the velocity semi-amplitude of the visible component. 
We first modelled the light curves with PHOEBE, for an estimation of the parameters $i$ and $P$. Having determined these values, we estimated the mass function from $K, P, G, i$ and found:

\begin{equation}
\frac{M^{3}_{1}}{(M_{1}+M_{2})^{2}} = 21.0 \pm\,3.4\,\emph{\msun}
\end{equation}

This large value implies a very large mass for the invisible component. If we further assume a mass of the Wolf Rayet
star $M_{2} = 15.3 \,\emph{\msun}$ as determined by \citet{Crowther06} for this system from the mass-luminosity relation of \citet{Schaer92} for hydrogen--free Wolf--Rayet stars, we end up with a third degree equation, the solution of which gives 
us the result of $\approxeq 40\,\emph{\msun}$ for the companion star.

\begin{table}[h]
\caption{Results for WR77o from combined $LC$ and $RV$ curve analysis}
\centering
\small\addtolength{\tabcolsep}{-3.0pt}
\begin{tabular}{l c }
\hline\hline
Parameter & Value \\
\hline\hline
Period, $P$ & 3.520 $\pm$ 0.003 $d$ \\
Time of primary eclipse, \emph{HJD$_0$} & 2453935.634 $\pm$ 0.003 \\
Inclination, $i$        & 65 $\pm$ 3 $\deg$\\
Surface potential, $\Omega_{1}$ & 3.47  $\pm$ 0.03  \\
Surface potential, $\Omega_{2}$ & 2.78 $\pm$ 0.03  \\
Light ratio in $I$, $L_{2}/L_{1}$ & 0.258 $\pm$ 0.005   \\
Light ratio in $R$, $L_{2}/L_{1}$ & 0.250 $\pm$ 0.005 \\
Semi-amplitude, $K_{2}$ & $360\pm30\; \rm km\; s^{-1}$ \\
Mass ratio, $q$ & 0.37 (fixed) \\
Systemic velocity, $\gamma$ & $-35\pm3\; \rm km\; s^{-1}$ \\
Semi-major axis, $a$ & $38 \pm2\; \emph{\rsun} $  \\
Radius$^{\dagger}$, $\rm r_{1,pole}$ & 0.32   $\pm$ 0.02  \\     
............ $\rm r_{1,point}$ & 0.33   $\pm$ 0.03   \\
............ $\rm r_{1,side}$ & 0.32   $\pm$ 0.02  \\
............ $\rm r_{1,back}$ & 0.33   $\pm$ 0.03  \\
............ $\rm r_{1}$$^{*}$ & 0.32  $\pm$ 0.03 \\
Radius$^{\dagger}$, $\rm r_{2,pole}$ & 0.24   $\pm$ 0.02  \\     
............ $\rm r_{2,point}$& 0.28   $\pm$ 0.03  \\
............ $\rm r_{2,side}$ & 0.25   $\pm$ 0.01  \\
............ $\rm r_{2,back}$ & 0.27   $\pm$ 0.02  \\
............ $\rm r_{2}$$^{*}$ & 0.25  $\pm$ 0.02  \\
\hline
Notes: \\
$\dagger$ The radii are given in units of \emph{a}. \\ $*$ Mean radius. 
\end{tabular}
\label{tab:wr77o}
\end{table}

We can continue with this
preliminary analysis to constrain 
some of the parameters and find a more robust 
solution using PHOEBE. Using the following values from \citet{Crowth07}: $\log(\rm L/\emph{\lsun})= 5.54$ and $T_\mathrm{eff}$ = 50\,000~K for a WN7 spectral type Wolf--Rayet star, we find the radius of the star to be $\sim 8$ $\,\emph{\rsun}$.

We ran PHOEBE using the $I$ and $R-$band light curves and the radial velocity curve of the secondary component, in the detached mode, for the following free parameters: \emph{i}, \emph{$a$}, \emph{$\gamma$}, \emph{$\Omega$}, \emph{$T_\mathrm{eff}$}, \emph{P}, \emph{HJD$_{0}$} and \emph{$L_1$}. 
The eccentricity was fixed to 0 and the mass ratio \emph{q} to 0.37, as determined above.

\begin{figure}[h]
\begin{center}$
\begin{array}{c}  
\includegraphics[scale=0.3]{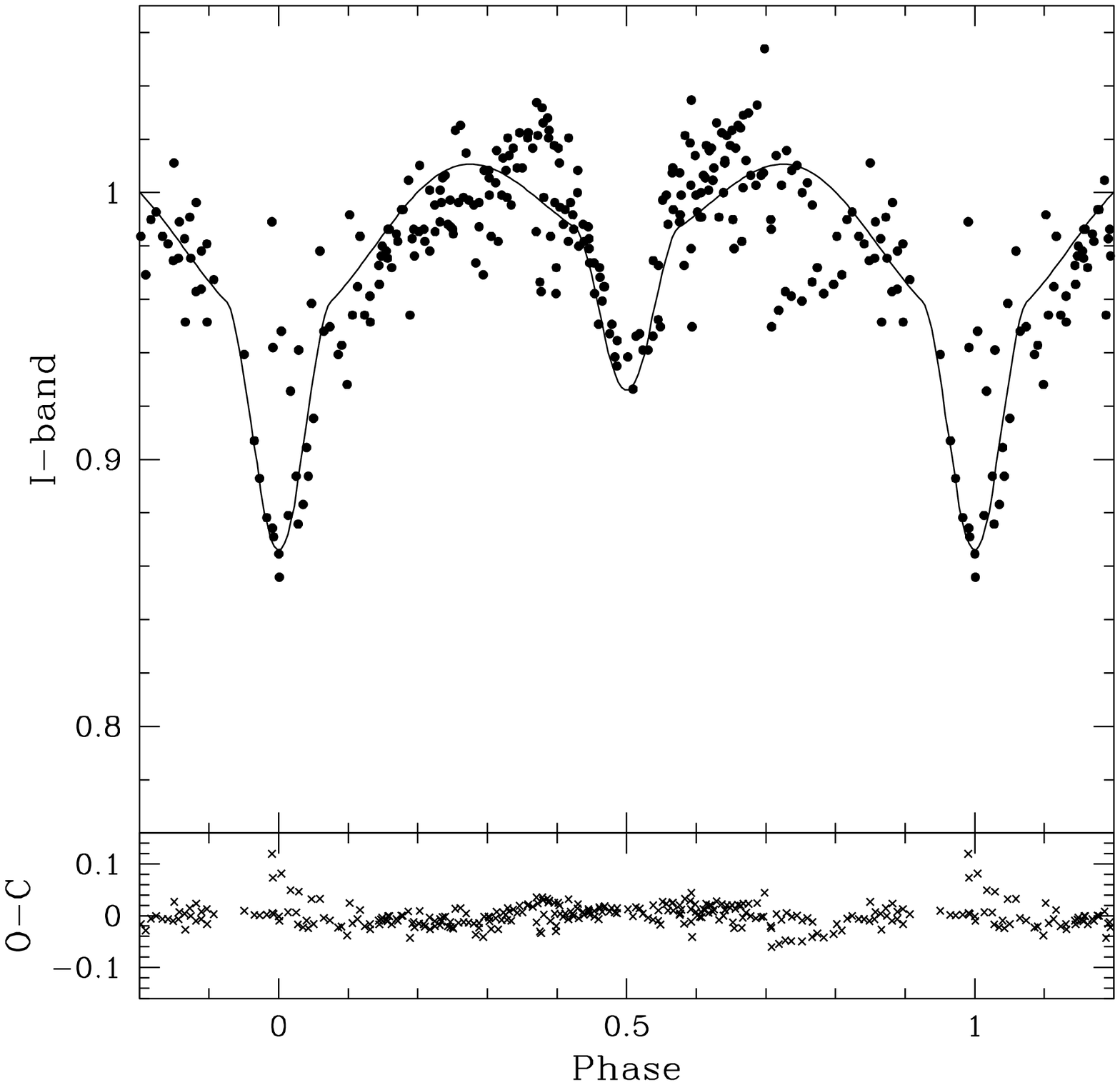} \\ \includegraphics[scale=0.3]{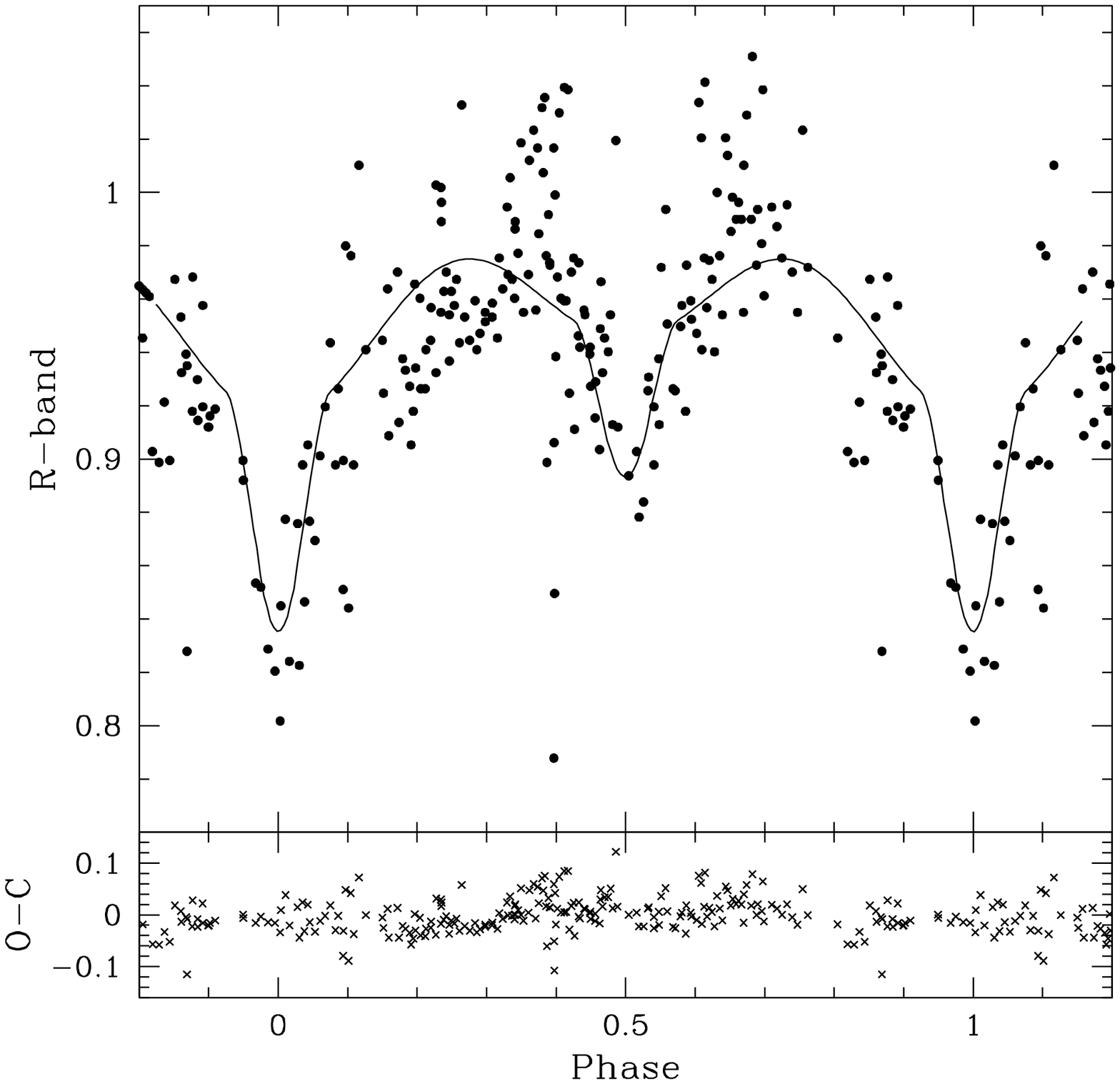} \\ \includegraphics[scale=0.3]{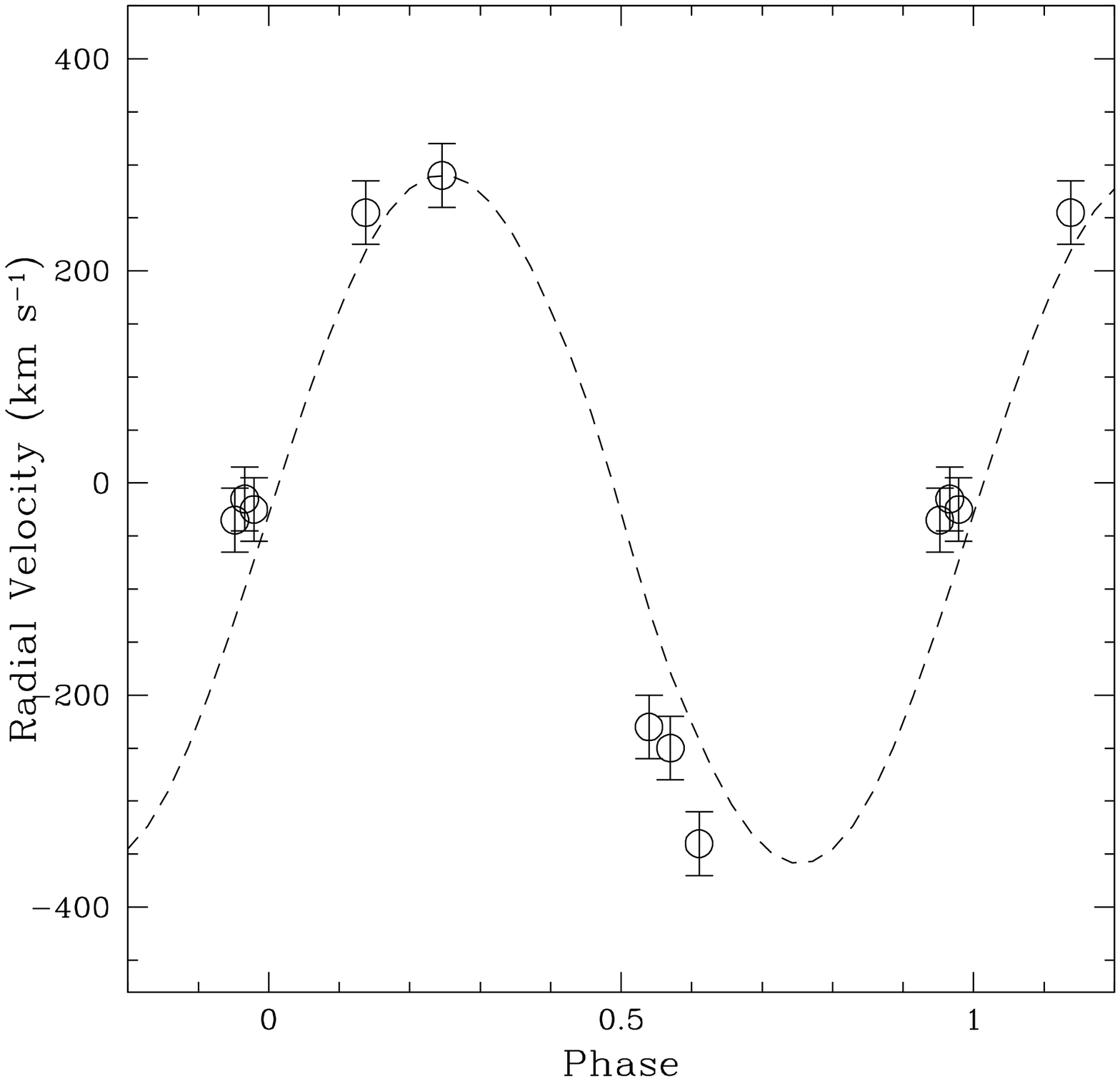}
\end{array}$
\end{center}
\caption{Phased light curves of WR77o in the $I$ and $R-$band in units of normalized light and radial velocity curve, with PHOEBE models overplotted (detached mode). The apparent magnitudes outside of eclipse are $I=14.3$ and $R=17.55$. The lower panel in each light curve shows the $O-C$ residuals.}
\label{fig:lcwr}
\end{figure}

Table~\ref{tab:wr77o} presents the parameters resulting from the
analysis with PHOEBE, which yielded a stable solution in agreement with our original estimates. The light and radial velocity
curves are shown in  Figure~\ref{fig:lcwr}. 

The Wolf-Rayet star was found to have 16$\,\emph{\msun}$ and 9.7$\,\emph{\rsun}$ 
while the primary companion 43$\,\emph{\msun}$ and 12.3$\,\emph{\rsun}$. The 
\emph{$T_\mathrm{eff}$} of the Wolf-Rayet star was found to be $44\,000~K$ which is quite close
to the expected one for its type. Even though the companion is characterized
by a higher {$T_\mathrm{eff}$} ($64\,000$~K) and a larger mass and radius, it is not visible in our spectra due to the low S/N, which adversely affects absorption 
lines. The high values of the filling factors, $F=0.75$ and $F=0.85$, respectively, indicate a post--mass transfer system.

Wolf--Rayet stars are intrinsically variable, therefore part of the observed scatter in the light curve is due to this effect. Modeling, which also takes into account atmospheric effects \citep[e.g.][]{Perrier09}, would possibly improve the fit and yield more accurate parameters. Furthermore, a more precise determination of the period would likely correct the slight phase shift of the radial velocity curve with respect to the model. We conclude that due to all the assumptions made for WR77o, the estimated parameters should be used with caution.

\begin{table}[h]
\caption{Physical parameters of the four Wd~1 eclipsing binaries}

\centering
\small\addtolength{\tabcolsep}{-3.0pt}
\begin{tabular}{c| c c c c }
\hline\hline
Parameter & W$_{DEB}$ & W36 & W13 & WR77o \\
\hline\hline
M$_1$ (M$_{\odot}$) & 14.8 $\pm$ 3.5 & 16.3 $\pm$ 1.5 & 23.1 $\pm$ 1.1 & 43.4 $\pm$ 6.8 \\
M$_2$ (M$_{\odot}$) & 11.9 $\pm$ 1.8 & 11.3 $\pm$ 1.8 & 32.9 $\pm$ 1.9 & 16.1 $\pm$ 2.5 \\
R$_1$ (R$_{\odot}$) & 6.9 $\pm$ 2.0 & 11.0 $\pm$ 1.2 & 23.0 $\pm$ 1.5 & 12.3 $\pm$ 2.0 \\
R$_2$ (R$_{\odot}$) & 5.3 $\pm$ 2.0 & 9.2 $\pm$ 1.2 & 21.3 $\pm$ 1.5 & 9.7 $\pm$ 2.0 \\
logg$_{1}$ & 3.93 $\pm$ 0.27 & 3.57 $\pm$ 0.10 & 3.08 $\pm$ 0.06 & 3.89 $\pm$ 0.16 \\
logg$_{2}$ & 4.07 $\pm$ 0.33 & 3.56 $\pm$ 0.13 & 3.30 $\pm$ 0.05 & 3.67 $\pm$ 0.19 \\
$T_\mathrm{eff_1}(\mathrm{K})$ & 29\,000 $\pm$ 580 & 30\,000 $\pm$ 1\,500 & 25\,000 (fixed) & 64\,000 $\pm$ 5\,000 \\
$T_\mathrm{eff_2}(\mathrm{K})$ & 27\,000 $\pm$ 570 & 25\,500 $\pm$ 2\,900 & 25\,000 $\pm$ 1\,000 & 44\,000 $\pm$ 4\,500 \\
log$L_{1}/L_{\odot}$ & 4.48 $\pm$ 0.25 & 5.19 $\pm$ 0.13 & 5.27 $\pm$ 0.06 & 6.36 $\pm$ 0.20 \\
log$L_{2}/L_{\odot}$ & 4.13 $\pm$ 0.33 & 4.51 $\pm$ 0.23 & 5.21 $\pm$ 0.09 & 5.50 $\pm$ 0.25 \\

\hline

\end{tabular}
\label{tab:bv}
\end{table}

\section{Evolutionary status} 

\begin{figure}[h]  
\includegraphics[scale=0.47]{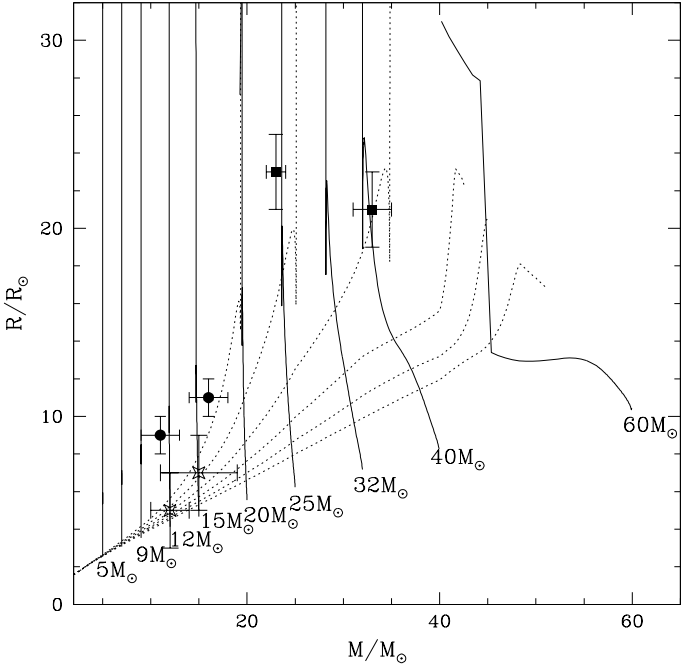}
\caption{Comparison of the parameters of W$_{DEB}$ (stars), W36 (circles) 
and W13 (squares) with evolutionary
tracks (solid lines) for stars with $5-60$ $\emph{\msun}$ and isochrones (dotted lines) for single rotating stars at solar metallicity from \citet{Ekstrom12}. Isochrones, from the bottom up, correspond to 3, 4, 5, 6, 8 and 10 Myr.
Single star isochrones are not compatible with the measured parameters for W36 and W13, indicating mass transfer.}
\label{fig:evol}
\end{figure}

\begin{figure}[h]  
\includegraphics[scale=0.47]{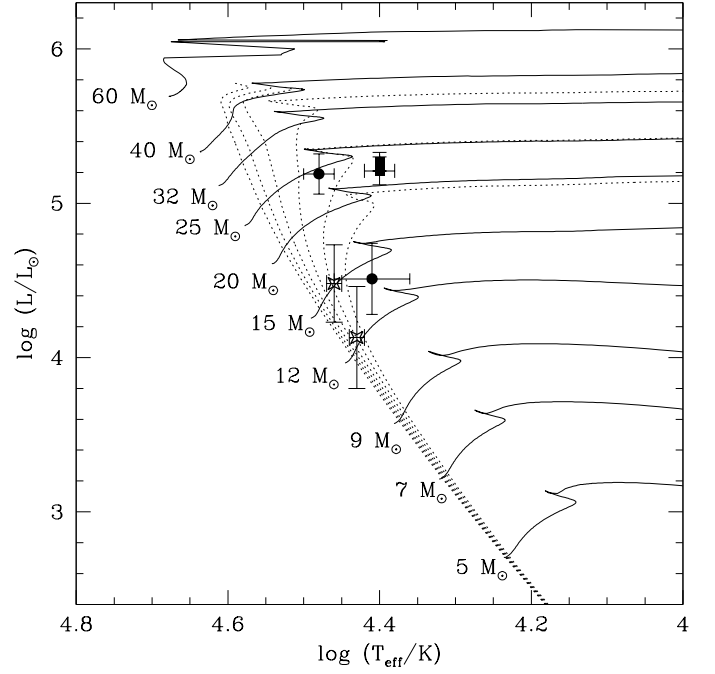}
\caption{Comparison of the parameters of W$_{DEB}$ (stars), W36 (circles) 
and W13 (squares) with evolutionary
tracks (solid lines) and isochrones (dotted lines) for single rotating stars at solar metallicity from \citet{Ekstrom12}. Isochrones, from left to right, correspond to 3, 4, 5, 6, 8 and 10 Myr. The comparison finds W$_{DEB}$ in good agreement with single star models, the components of W36 to appear overluminous for their mass, and the primary of W13 in agreement, while the secondary appears underluminous.}
\label{fig:hr}
\end{figure}

For a comparison of our measured parameters with evolutionary models, we
used the models of \citet{Ekstrom12}, for single stars in the range $0.8-120\,\emph{\msun}$ at
solar metallicity (Z=0.014), which include rotation. These evolutionary models are not appropriate for post mass-transfer binaries, however we proceed with the comparison, as it provides useful information, while computing binary evolution models is outside the scope of this work. Figure~\ref{fig:evol} compares the parameters of each binary (masses vs. radii) with evolutionary
tracks and isochrones for single rotating stars at solar metallicity, while Figure~\ref{fig:hr} plots them on an H-R diagram.  

The position of the components of W36 on the mass-radius diagram (Figure~\ref{fig:evol}) indicates an age of 
$>$ 10 Myr for the secondary, while the discrepancy is smaller in the H-R diagram. The apparent conflict with previous studies that find an age of 4.5--5 Myr \citep{Crowther06} can be resolved by considering that W36 is a contact binary, which has undergone mass transfer. On the other hand, W$_{DEB}$ with its unevolved members of 15$\,\emph{\msun}$ and 12$\,\emph{\msun}$ is, 
in principle, ideal for estimating the age of the cluster. Its position on this diagram
supports an age that is consistent within errors to previous studies, as detached unevolved systems can be treated as single stars. The position of the secondary of W13 also indicates a similar age for Wd~1. WR77o is not plotted in the diagram since we were unable to detect both stars spectroscopically to determine accurate parameters.  

The position of the components of W36 on the H-R diagram (Figure~\ref{fig:hr}) show that both stars appear overluminous for their mass, which provides further evidence that the system has undergone mass transfer. \citet{North10} have also reported instances of non--detached eclipsing systems whose components appear overluminous or underluminous for their mass when compared with tracks for single stars. The position of W$_{DEB}$ is in good agreement with the single star models although the current error bars do not allow a precise age estimation. Finally, the primary of W13 is in agreement with the models, while the secondary appears underluminous, once again providing evidence for mass transfer.

\vspace{0.5cm}

\section{Distance to Westerlund~1} 

The distance to Westerlund~1 has previously been estimated to be $\sim4-5$
kpc, using a variety of methods, e.g.: WR stars
\citep[$5.0^{+0.5}_{-1.0}$ kpc;][]{Crowther06}, \ion{H}{I} observations
\citep[$3.9 \pm 0.7$ kpc;][]{Kothes07}, OB supergiants \citep[$>5$
  kpc;][]{Negueruela10}, and pre-main sequence stars \citep[$4.0\pm0.2$
  kpc;][]{Gennaro11}. Eclipsing binaries provide powerful tools for
measuring independent and accurate distances to stellar clusters
\citep[e.g.][]{Rauw07} and nearby galaxies \citep[e.g.][]{Bonanos06,
  Bonanos11}. The analysis of the four systems described in the previous
sections therefore opens the way for direct, independent eclipsing binary
distance measurements to Wd~1.


Of the four systems, W13 has the most accurately determined radii and
effective temperatures, and so we proceed to compute the distance to
Westerlund~1 based on this eclipsing binary. \citet{Ritchie10}
determined accurate spectral types for the components of W13: B0.5
Ia$^{+}$/WNVL for the emission line component and O9.5$-$B0.5I for the
absorption line component, which correspond to a $\rm
T_{eff}=25\,000\pm2\,000$~K. We adopted this conservative error for both components for the distance estimation. Using the radii resulting from our
analysis, we compute $\log(\rm L_{1}/\emph{\lsun})= 5.27\pm0.20\,$ and
$\log(\rm L_{2}/\emph{\lsun})= 5.20\pm0.20\,$, and for the
combined light: $\log(\rm L_{total}/\emph{\lsun})= 5.54\pm0.11$. The absolute
magnitude of a star at a certain wavelength $\lambda$ is given by:

\begin{equation}
M_{\lambda}=\emph{\msun}^{bol}-BC_{\lambda}-2.5 \log \left( \frac{\rm L}{\emph{\lsun}} \right)
\end{equation}

\noindent where $\emph{\msun}^{bol}=4.75$~mag is the bolometric luminosity of
the Sun and $BC_{\lambda}$ the bolometric correction. We used theoretical \emph{BC} values
from \citet{Martins06}, which are based on state-of-the-art CMFGEN model
atmospheres \citep{Hillier98}. For an O9.5I star, the theoretical
BC$_V=-2.62$~mag, which yields $M_V=-6.47$~mag.

The optical out-of-eclipse magnitudes of W13 are $V=17.50$, $R=14.66$ and
$I=12.02$ \citep{Bonanos07}. \citet{Clark10} found the $V-$band
calibration of \citet{Bonanos07} to be offset to that of
\citet{Clark05}, while the $R$ and $I-$band calibrations were in good
agreement, and reported magnitudes for W13 of $B=21.1,\, V=17.19,\,
R=14.63, \,I=12.06$, corresponding to phase $\sim0.25$. For the distance
estimation, we adopt the higher-quality photometry of \citet{Clark05},
obtained with a larger telescope and a conservative error of 0.04~mag in each band,
which corresponds to the difference in the $R$ and $I-$band
measurements.

The distance $d$ is estimated from the distance modulus equation:

\begin{equation}
m_{v} - M_{V} - A_{V}= 5 \times \log \left( \frac{d}{10pc}\right)
\label{distmod}
\end{equation}
\noindent where $m_{v}$ is the apparent $V-$band magnitude, and
$A_{V}=R_{V}\times E(B-V)$ is the extinction. From the photometry of
\citet{Clark05} and adopting $(B-V)_{0}=-0.26$ from \citet{Martins06}, we obtain $E(B-V)=4.17\pm0.05$~mag,
which yields $A_{V}=12.93\pm2.09$~mag (for $R_{V}=3.1\pm0.5$), in
agreement with the average value found of $A_{V}=13.6$~mag found by
\citet{Clark05} for OB supergiants. Equation~\ref{distmod} therefore
yields $d=1.4^{+2.3}_{-0.9}$~kpc, which is a very imprecise estimate due
to the large uncertainty in $A_{V}$, demonstrating that near--infrared photometry and an accurate determination of the extinction is essential for an accurate distance measurement.

%
%

We therefore proceed to use near--infrared photometry, which is less affected by dust extinction. The available 2MASS photometry of W13 (at an unknown phase) is $J=9.051\pm0.020$ \footnote{The $H$ and $K$ bands only have upper limits.}, and we adopt a conservative error of 0.16~mag (the depth of W13 primary eclipse), i.e. $J=9.05\pm0.16$~mag.

For an O9.5I star, the theoretical BC$_{J}$$=-3.24$~mag, which yields $M_{J}=-5.85$~mag. We use the following equation to estimate the distance:

\begin{equation}
m_{J} - M_{J} - A_{J}= 5 \times \log \left( \frac{d}{10pc}\right)
\label{distmodJ}
\end{equation}

We adopt $A_{J}/A_{K}=2.50\pm0.15$ \citep{Indebetouw05} and the
average value for $A_{K}=0.82$~mag, determined toward the WR star
``R'' (or WR77q), a close neighbor of W13 $\sim6\arcsec$ away by
\citet{Crowther06}. We note that Star ``U'' (or WR77s), also nearby, has
$A_{K}=0.74$~mag, therefore we adopt a conservative error in $A_{K}$ of 0.1~mag. We find $A_{J}=2.05\pm 0.28$~mag and derive a distance: $d=3.71\pm0.55$~kpc to W13, which is the first eclipsing binary distance to the cluster. This value with an accuracy of 15\% is in
good agreement with recent estimates, which is noteworthy as eclipsing binaries are reliable, direct and independent distance indicators. For future determinations of the true distance modulus ($DM_{0}$) with better estimates of the extinction, we also report the value $DM_{0}+A_{J}=14.90\pm0.16$~mag. The precision of our measurement can be improved to $<5\%$ by:
(a) future near--infrared photometry to unambiguously disentangle the
extinction and reddening law toward W13, and (b) spectral energy
distribution (SED) modeling to improve the temperature determination.

\section{Discussion and conclusions} 

We have presented accurate (better than 15\%) fundamental parameters of
four massive eclipsing binaries in the young massive cluster Westerlund~1, 
confirming they are all cluster members and demonstrating once more
the success of the survey strategy. The mass range found for the eight
component stars is $11-43\,\emph{\msun}$, and three of the systems (W13, W36
and WR77o) seem to be post mass-transfer binaries. Our results for W13 are in
agreement with those presented in the independent study by
\citet{Ritchie10}. We were able to improve the accuracy of the
parameters of W13 and therefore provide confirmation on the dynamical constraint
on the high progenitor mass ($>40\,\emph{\msun}$) of the magnetar CSO
J164710.2$-$455216 \citep{Muno06} reported by \citet{Ritchie10}, which accounted for binary evolution. Such massive stars are believed to form black holes rather than neutron stars, at least in a single star scenario. However, a strong interaction, as that of mass stripping in a binary system, can change the progenitor mass range significantly, without causing any problem in our understanding of stellar evolution \citep{Belczynski08}. Our
discovery of a massive $\sim40\,\emph{\msun}$ component for WR77o is also
significant as it will serve as a second dynamical constraint on the
mass of the progenitor of the magnetar. Future high S/N spectra should
reveal the massive component, which, combined with a new, higher quality light curve,
will enable an accurate determination of the component masses.

\begin{figure}[h]  
\includegraphics[scale=0.45]{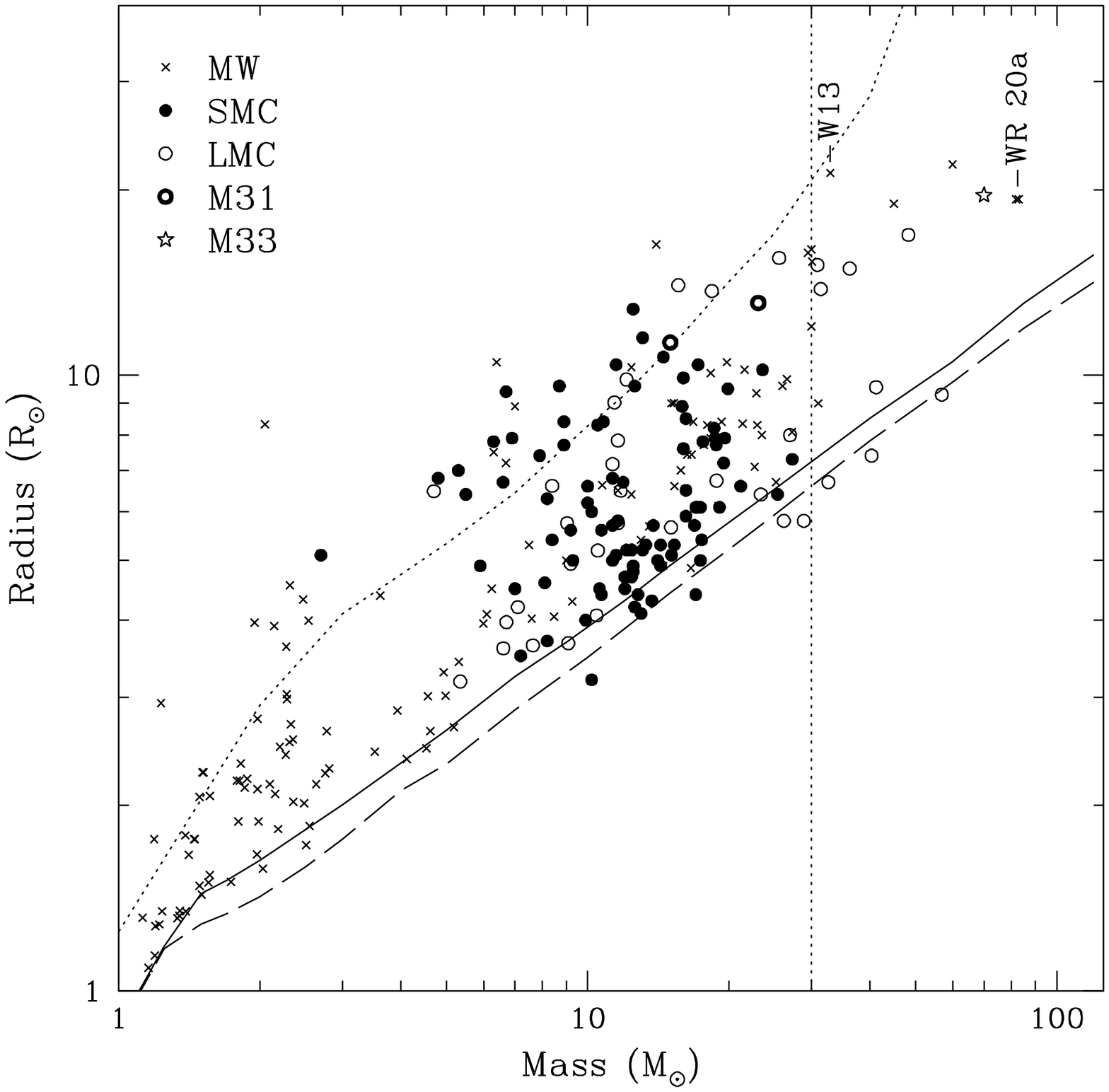}
\caption{Mass and radius determinations of stars in eclipsing binaries, accurate
to $\leq 10 \%$ and complete $\geq 30 \emph{\msun}$ compiled from the literature by \citet{Bonanos09}, now updated with the massive secondary of W13. The solid line is the Z = 0.02 zero--age main sequence (ZAMS) from \citet{Schall92}; the dashed line is the Z = 0.008 ZAMS from \citet{Schaer93}; the dotted line is the terminal age main sequence (TAMS). W13 brings the total number of stars fulfilling the requirements of mass and accuracy to 18.}
\label{fig:comp}
\end{figure}

W13 is also noteworthy because the absorption component
($33\pm2\,\emph{\msun}$) fulfills the criteria set forth by \citet{Bonanos09}
and increases the total number of very massive stars ($>30\,\emph{\msun}$) with
$<10\%$ measurements from eclipsing binaries to 18. Figure~\ref{fig:comp} shows the updated mass radius diagram with eclipsing binary measurements accurate to 10\% and complete $>30\,\emph{\msun}$. The slow rate of
increase of this sample (one star per year) is a reflection of the difficulty of
identifying very massive stars and obtaining follow-up observations. The
availability of eclipsing binaries with accurate parameters now enables direct, independent eclipsing binary distance
determinations to the cluster. We have used our results for W13 to
estimate the first eclipsing binary distance to the cluster and therefore the magnetar, finding a
value of $d=3.7\pm0.6$~kpc, in agreement with previous determinations using
other methods. Future near-infrared photometry, a better light curve, in
addition to careful modelling of the SED has the potential to provide a
measurement accurate to $<5\%$, which will be crucial for our comprehension of magnetars.

Lastly, we find $W_{DEB}$, despite its distance from the core of the cluster, to have
the same systemic velocity as the other eclipsing systems, indicating it
was ejected due south. Future proper motion studies will determine the
time of ejection and possible connection with other massive cluster
stars located outside the core, such as the magnetar, luminous blue
variable, WR77aa (T) and WR77b \citep[N; see][]{Crowther06}, possibly
indicating frequent ejections of stars from the cluster core, as was
reported for Westerlund~2 \citep{Roman-Lopes11}.

In conclusion, we note that young massive clusters remain largely
unexplored regarding their binary star content, although binaries are
extremely useful tools for providing constraints for models of
evolution and formation for single and binary stars, at a range of
metallicities. In particular for Wd~1, we have used the four known eclipsing
binaries to obtain fundamental parameters for eight massive stars and to derive the first 
independent distance determination to the cluster.

\begin{acknowledgements}
The authors thank Rodolfo Barba for obtaining the IMACS observations, 
Jose L. Prieto for obtaining a spectrum of W$_{DEB}$ with Magellan in
September 2011.
We also thank the referee, Gregor Rauw, for a careful reading of the manuscript and for providing helpful comments and suggestions that improved the paper. AZB acknowledges research and travel support from the
Carnegie Institution of Washington through a Vera Rubin Fellowship in
2005-2008. EK and AZB acknowledge research and travel support from the
European Commission Framework Program Seven under the Marie Curie
International Reintegration Grant PIRG04-GA-2008-239335.
\end{acknowledgements}

\bibliographystyle{aa} 
\bibliography{ref}

\begin{thebibliography}{44}
\expandafter\ifx\csname natexlab\endcsname\relax\def\natexlab#1{#1}\fi

\bibitem[{{Andersen}(1991)}]{Andersen91}
{Andersen}, J. 1991, \aapr, 3, 91

\bibitem[{{Belczynski} \& {Taam}(2008)}]{Belczynski08}
{Belczynski}, K. \& {Taam}, R.~E. 2008, \apj, 685, 400

\bibitem[{{Bernstein} {et~al.}(2003)}]{Bernstein03}
{Bernstein}, R. {et~al.} 2003, in SPIE, Vol. 4841, pp. 1694-1704, ed. M.~{Iye}
  \& A.~F.~M. {Moorwood}

\bibitem[{{Bonanos}(2007)}]{Bonanos07}
{Bonanos}, A.~Z. 2007, \aj, 133, 2696

\bibitem[{{Bonanos}(2009)}]{Bonanos09}
{Bonanos}, A.~Z. 2009, \apj, 691, 407

\bibitem[{{Bonanos}(2010)}]{Bonanos10}
{Bonanos}, A.~Z. 2010, in ASP Conf.\ Ser., Vol. 425, Hot and Cool: Bridging
  Gaps in Massive Star Evolution, ed. {C.~Leitherer, P.~Bennett, P.~Morris, \&
  J.~van Loon} (San Francisco: ASP), 31

\bibitem[{{Bonanos} {et~al.}(2011){Bonanos}, {Castro}, {Macri},
  {et~al.}}]{Bonanos11}
{Bonanos}, A.~Z., {Castro}, N., {Macri}, L.~M., {et~al.} 2011, \apjl, 729, L9

\bibitem[{{Bonanos} {et~al.}(2006){Bonanos}, {Stanek}, {Kudritzki},
  {et~al.}}]{Bonanos06}
{Bonanos}, A.~Z., {Stanek}, K.~Z., {Kudritzki}, R.~P., {et~al.} 2006, \apj,
  652, 313

\bibitem[{{Clark} \& {Negueruela}(2004)}]{Clar04}
{Clark}, J.~S. \& {Negueruela}, I. 2004, \aap, 413, L15

\bibitem[{{Clark} {et~al.}(2005){Clark}, {Negueruela}, {Crowther},
  {et~al.}}]{Clark05}
{Clark}, J.~S., {Negueruela}, I., {Crowther}, P.~A., {et~al.} 2005, \aap, 434,
  949

\bibitem[{{Clark} {et~al.}(2010){Clark}, {Ritchie}, \& {Negueruela}}]{Clark10}
{Clark}, J.~S., {Ritchie}, B.~W., \& {Negueruela}, I. 2010, \aap, 514, A87

\bibitem[{{Crowther}(2007)}]{Crowth07}
{Crowther}, P.~A. 2007, \araa, 45, 177

\bibitem[{{Crowther} {et~al.}(2006){Crowther}, {Hadfield}, {Clark},
  {et~al.}}]{Crowther06}
{Crowther}, P.~A., {Hadfield}, L.~J., {Clark}, J.~S., {et~al.} 2006, \mnras,
  1067

\bibitem[{{Dressler} {et~al.}(2011){Dressler}, {Bigelow}, {Hare},
  {et~al.}}]{Dressler11}
{Dressler}, A., {Bigelow}, B., {Hare}, T., {et~al.} 2011, \pasp, 123, 288

\bibitem[{{Ekstr{\"o}m} {et~al.}(2012){Ekstr{\"o}m}, {Georgy}, {Eggenberger},
  {Meynet}, {Mowlavi}, {Wyttenbach}, {Granada}, {Decressin}, {Hirschi},
  {Frischknecht}, {Charbonnel}, \& {Maeder}}]{Ekstrom12}
{Ekstr{\"o}m}, S., {Georgy}, C., {Eggenberger}, P., {et~al.} 2012, \aap, 537,
  A146

\bibitem[{{Gennaro} {et~al.}(2011){Gennaro}, {Brandner}, {Stolte}, \&
  {Henning}}]{Gennaro11}
{Gennaro}, M., {Brandner}, W., {Stolte}, A., \& {Henning}, T. 2011, \mnras,
  412, 2469

\bibitem[{{Gr{\"a}fener} {et~al.}(2002){Gr{\"a}fener}, {Koesterke}, \&
  {Hamann}}]{Graefener02}
{Gr{\"a}fener}, G., {Koesterke}, L., \& {Hamann}, W.-R. 2002, \aap, 387, 244

\bibitem[{{Hamann} \& {Gr{\"a}fener}(2003)}]{Hamann03}
{Hamann}, W.-R. \& {Gr{\"a}fener}, G. 2003, \aap, 410, 993

\bibitem[{{Hamann} \& {Gr{\"a}fener}(2004)}]{Hamann04}
{Hamann}, W.-R. \& {Gr{\"a}fener}, G. 2004, \aap, 427, 697

\bibitem[{{Hillier} \& {Miller}(1998)}]{Hillier98}
{Hillier}, D.~J. \& {Miller}, D.~L. 1998, \apj, 496, 407

\bibitem[{{Indebetouw} {et~al.}(2005){Indebetouw}, {Mathis}, {Babler},
  {et~al.}}]{Indebetouw05}
{Indebetouw}, R., {Mathis}, J.~S., {Babler}, B.~L., {et~al.} 2005, \apj, 619,
  931

\bibitem[{{Kelson}(2003)}]{Kelson03}
{Kelson}, D.~D. 2003, \pasp, 115, 688

\bibitem[{{Kelson} {et~al.}(2000)}]{Kelson00}
{Kelson}, D.~D. {et~al.} 2000, \apj, 531, 159

\bibitem[{{Kothes} \& {Dougherty}(2007)}]{Kothes07}
{Kothes}, R. \& {Dougherty}, S.~M. 2007, \aap, 468, 993

\bibitem[{{Lanz} \& {Hubeny}(2003)}]{Lanz03}
{Lanz}, T. \& {Hubeny}, I. 2003, \apjs, 146, 417

\bibitem[{{Lanz} \& {Hubeny}(2007)}]{Lanz07}
{Lanz}, T. \& {Hubeny}, I. 2007, \apjs, 169, 83

\bibitem[{{Martins} {et~al.}(2006){Martins}, {Trippe}, {Paumard},
  {et~al.}}]{Martins06}
{Martins}, F., {Trippe}, S., {Paumard}, T., {et~al.} 2006, \apjl, 649, L103

\bibitem[{{Massey} \& {Conti}(1977)}]{Massey77}
{Massey}, P. \& {Conti}, P.~S. 1977, \apj, 218, 431

\bibitem[{{Muno} {et~al.}(2006){Muno}, {Clark}, {Crowther}, {et~al.}}]{Muno06}
{Muno}, M.~P., {Clark}, J.~S., {Crowther}, P.~A., {et~al.} 2006, \apjl, 636,
  L41

\bibitem[{{Negueruela} \& {Clark}(2005)}]{Negueruela05}
{Negueruela}, I. \& {Clark}, J.~S. 2005, \aap, 436, 541

\bibitem[{{Negueruela} {et~al.}(2010){Negueruela}, {Clark}, \&
  {Ritchie}}]{Negueruela10}
{Negueruela}, I., {Clark}, J.~S., \& {Ritchie}, B.~W. 2010, \aap, 516, A78

\bibitem[{{North} {et~al.}(2010){North}, {Gauderon}, {Barblan}, \&
  {Royer}}]{North10}
{North}, P., {Gauderon}, R., {Barblan}, F., \& {Royer}, F. 2010, \aap, 520, A74

\bibitem[{{Perrier} {et~al.}(2009){Perrier}, {Breysacher}, \&
  {Rauw}}]{Perrier09}
{Perrier}, C., {Breysacher}, J., \& {Rauw}, G. 2009, \aap, 503, 963

\bibitem[{{Pr{\v s}a} \& {Zwitter}(2005)}]{Prsa05}
{Pr{\v s}a}, A. \& {Zwitter}, T. 2005, \apj, 628, 426

\bibitem[{{Rauw} {et~al.}(2007){Rauw}, {Manfroid}, {Gosset}, {et~al.}}]{Rauw07}
{Rauw}, G., {Manfroid}, J., {Gosset}, E., {et~al.} 2007, \aap, 463, 981

\bibitem[{{Ritchie} {et~al.}(2010){Ritchie}, {Clark}, {Negueruela},
  {et~al.}}]{Ritchie10}
{Ritchie}, B.~W., {Clark}, J.~S., {Negueruela}, I., {et~al.} 2010, \aap, 520,
  A48

\bibitem[{{Roman-Lopes} {et~al.}(2011){Roman-Lopes}, {Barba}, \&
  {Morrell}}]{Roman-Lopes11}
{Roman-Lopes}, A., {Barba}, R.~H., \& {Morrell}, N.~I. 2011, \mnras, 416, 501

\bibitem[{{Schaerer} {et~al.}(1993){Schaerer}, {Charbonnel}, {Meynet},
  {Maeder}, \& {Schaller}}]{Schaer93}
{Schaerer}, D., {Charbonnel}, C., {Meynet}, G., {Maeder}, A., \& {Schaller}, G.
  1993, \aaps, 102, 339

\bibitem[{{Schaerer} \& {Maeder}(1992)}]{Schaer92}
{Schaerer}, D. \& {Maeder}, A. 1992, \aap, 263, 129

\bibitem[{{Schaller} {et~al.}(1992){Schaller}, {Schaerer}, {Meynet}, \&
  {Maeder}}]{Schall92}
{Schaller}, G., {Schaerer}, D., {Meynet}, G., \& {Maeder}, A. 1992, \aaps, 96,
  269

\bibitem[{{Schmutz} {et~al.}(1992){Schmutz}, {Leitherer}, \&
  {Gruenwald}}]{Schmutz92}
{Schmutz}, W., {Leitherer}, C., \& {Gruenwald}, R. 1992, \pasp, 104, 1164

\bibitem[{{Stroud} {et~al.}(2010){Stroud}, {Clark}, {Negueruela},
  {et~al.}}]{Stroud10}
{Stroud}, V.~E., {Clark}, J.~S., {Negueruela}, I., {et~al.} 2010, \aap, 511,
  A84

\bibitem[{{Torres} {et~al.}(2010){Torres}, {Andersen}, \&
  {Gim{\'e}nez}}]{Torres10}
{Torres}, G., {Andersen}, J., \& {Gim{\'e}nez}, A. 2010, \aapr, 18, 67

\bibitem[{{Wilson} \& {Devinney}(1971)}]{Wilson71}
{Wilson}, R.~E. \& {Devinney}, E.~J. 1971, \apj, 166, 605

\end{thebibliography}

\end{document}